\documentclass[times,10pt,twocolumn]{article}
\usepackage{latex8}
\usepackage{times}
\usepackage[latin1]{inputenc}

\usepackage{epsfig}
\usepackage{latexsym}
\usepackage{amsmath}
\usepackage{amssymb}
\usepackage{epsfig}

\usepackage{subfigure}

\def\punto{$\hspace*{\fill}\Box$}
\newcommand{\nop}[1]{}

\newtheorem{theorem}{Theorem}[section]

\newtheorem{example}[theorem]{Example}
\newtheorem{definition}[theorem]{Definition}
\newtheorem{proposition}[theorem]{Proposition}

\newtheorem{corollary}[theorem]{Corollary}
\newtheorem{lemma}[theorem]{Lemma}
\newtheorem{remark}[theorem]{Remark}

\title{{\LARGE {10\textasciicircum(10\textasciicircum 6)}} Worlds and Beyond:
Efficient Representation and Processing of Incomplete Information}

\author{
	Lyublena Antova\\
	Cornell University\\
	lantova@cs.cornell.edu	
	\and
	Christoph Koch\\
	Cornell University\\
	koch@cs.cornell.edu
	\and
	Dan Olteanu\\
	Oxford University\\
  dan.olteanu@comlab.ox.ac.uk
}
\begin{document}

\maketitle

\begin{abstract}
%Current systems and formalisms for representing incomplete information
%generally suffer from at least one of two weaknesses. Either they are not
%strong enough for representing results of simple queries, or the handling and
%processing of the data, e.g. for query evaluation, is intractable. 

%In this paper, we present a decomposition-based approach to addressing this
%problem. We introduce {\em world-set decompositions (WSDs)}\/, a
%space-efficient formalism for representing any finite set of possible worlds
%over relational databases. WSDs are therefore a strong representation system
%for any relational query language. We study the problem of efficiently
%evaluating relational algebra queries on sets of worlds represented by WSDs.
%We also evaluate our technique experimentally in a large census data scenario
%and show that it is both scalable and efficient.
  
  We present a decomposition-based approach to managing incomplete
  information. We introduce {\em world-set decompositions (WSDs)}\/, a
  space-efficient and complete representation system for finite sets
  of worlds. We study the problem of efficiently evaluating relational
  algebra queries on world-sets represented by WSDs.  We also evaluate
  our technique experimentally in a large census data scenario and
  show that it is both scalable and efficient.
  
%  \keywords{Incomplete information \and Uncertain and probabilistic databases \and Query processing}
\end{abstract}
%\vspace*{-1em}

\section{Introduction}
\label{sec:introduction}

% \vspace{-3mm}

Incomplete information is commonplace in real-world databases.  Classical
examples can be found in data integration and wrapping applications,
linguistic collections, or whenever information is manually entered and is
therefore prone to inaccuracy or partiality.

There has been little research so far into expressive {\em yet scalable}\/
systems for representing incomplete information.  Current techniques can be
classified into two groups. The first group includes representation systems
such as {\em v-tables} \cite{IL1984} and {\em or-set relations} \cite{INV1991}
which are not strong enough to represent the results of relational algebra
queries within the same formalism.  In v-tables the tuples can contain both
constants and variables, and each combination of possible values for the
variables yields a possible world. Relations with or-sets can be viewed as
v-tables, where each variable occurs only at a single position in the table
and can only take values from a fixed finite set, the or-set of the field
occupied by the variable. The so-called {\em c-tables}~\cite{IL1984} belong to
the second group of formalisms.  They extend v-tables with conditions
specified by logical formulas over the variables, thus constraining the
possible values.  Although c-tables are a strong representation system, they
have not found application in practice.
The main reason for this is probably that managing c-tables directly is
rather inefficient. Even very basic problems such as deciding whether a tuple
is in at least one world represented by the c-table are
intractable~\cite{AKG1991}.

\begin{figure}
        \begin{center}
                \epsfig{file=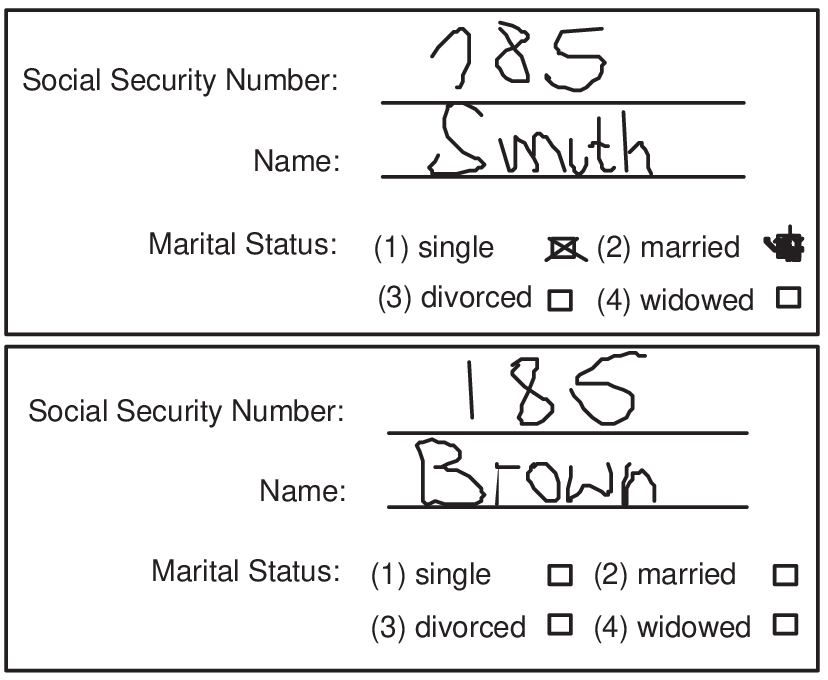, width=.4\textwidth}
        \end{center}
% \vspace{-6mm}
        \caption{Two completed survey forms.}
        \label{fig:census}
\end{figure}

As a motivation, consider two manually completed forms that may
originate from a census
% or some other kind of survey 
and which allow for more than one interpretation
(Figure~\ref{fig:census}). For simplicity we assume that social
security numbers consist of only three digits. For instance, Smith's
social security number can be read either as ``185'' or as ``785''.
We can represent the available information using a relation with
or-sets:
% \vspace{-0.1cm}
{\footnotesize
\begin{center}
\begin{tabular}{c|ccc}
(TID) & S & N & M \\
\hline
$t_1$ & \{ 185, 785 \} & Smith & \{ 1, 2 \} \\
$t_2$ & \{ 185, 186 \} & Brown & \{ 1, 2, 3, 4 \} \\
\end{tabular}
\end{center}
}
% \vspace{-0.1cm}

It is easy to see that this or-set relation represents
$2\cdot2\cdot2\cdot4 = 32$ {\em possible worlds}\/.

Given such an incompletely specified database, it must of course be possible
to access and process the data.  Two data management tasks shall be pointed
out as particularly important, the evaluation of queries on the data and {\em
  data cleaning}~\cite{RD2000,GFSS2000,RH2001}, by which certain worlds can be
shown to be impossible and can be excluded.  The results of both types of
operation turn out not to be representable by or-set relations in general.
Consider for example the integrity constraint that all social security numbers
be unique.  For our example database, this constraint excludes 8 of the 32
worlds, namely those in which both tuples have the value 185 as social
security number.  It is impossible to represent the remaining 24 worlds using
or-set relations.  This is an example of a constraint that can be used for
data cleaning; similar problems are observed with queries, e.g., the query
asking for pairs of persons with differing social security numbers.

What we could do is store each world
explicitly using a table called a {\em
world-set relation}\/ of a given set of worlds. Each tuple in
this table represents one world and is the concatenation of
all tuples in that world (see Figure \ref{fig:bigtable}).

The most striking problem of world-set relations is their size. If we
conduct a survey of 50 questions on a population of 200 million and we
assume that one in $10^4$ answers can be read in just two different
ways, we get $2^{10^6}$ worlds. Each such world is a substantial table
of 50 columns and $2\cdot 10^8$ rows.  We cannot store all these
worlds explicitly in a world-set relation (which would have $10^{10}$
columns and $2^{10^6}$ rows). Data cleaning will often eliminate only
some of these worlds, so a DBMS should manage those that remain.

\begin{figure}[tb]
{\footnotesize
\begin{center}
\begin{tabular}{cccccc}
$t_1$.S & $t_1$.N & $t_1$.M &
$t_2$.S & $t_2$.N & $t_2$.M \\
\hline
185 & Smith & 1 & 186 & Brown & 1 \\
185 & Smith & 1 & 186 & Brown & 2 \\
185 & Smith & 1 & 186 & Brown & 3 \\
185 & Smith & 1 & 186 & Brown & 4 \\
185 & Smith & 2 & 186 & Brown & 1 \\
\multicolumn{6}{c}{\vdots}\\
\nop{ % removed for ICDE
185 & Smith & 2 & 186 & Brown & 2 \\
185 & Smith & 2 & 186 & Brown & 3 \\
185 & Smith & 2 & 186 & Brown & 4 \\
785 & Smith & 1 & 185 & Brown & 1 \\
785 & Smith & 1 & 185 & Brown & 2 \\
785 & Smith & 1 & 185 & Brown & 3 \\
785 & Smith & 1 & 185 & Brown & 4 \\
785 & Smith & 1 & 186 & Brown & 1 \\
785 & Smith & 1 & 186 & Brown & 2 \\
785 & Smith & 1 & 186 & Brown & 3 \\
785 & Smith & 1 & 186 & Brown & 4 \\
785 & Smith & 2 & 185 & Brown & 1 \\
785 & Smith & 2 & 185 & Brown & 2 \\
785 & Smith & 2 & 185 & Brown & 3 \\
785 & Smith & 2 & 185 & Brown & 4 \\
785 & Smith & 2 & 186 & Brown & 1 \\
785 & Smith & 2 & 186 & Brown & 2 \\
785 & Smith & 2 & 186 & Brown & 3 \\
}
785 & Smith & 2 & 186 & Brown & 4 \\
\hline
\end{tabular}
\end{center}
}
% \vspace{-4mm}
\caption{World-set relation containing only 
  worlds with unique social security numbers.}  % \vspace{-5mm}
%\vspace*{-1em}
\label{fig:bigtable}
\end{figure}

This article aims at dealing with this complexity and proposes the new
notion of {\em world-set decompositions (WSDs)}\/.  These are
decompositions of a world-set relation into several relations such
that their product (using the product operation of relational algebra)
is again the world-set relation.

\begin{example}
 \em
The world-set represented by our initial or-set
relation can also be represented by the product
\end{example}
%
% in Figure \ref{fig:initialdec}.
%
%\begin{figure}
% \vspace{-2mm}
{\footnotesize
\[
\hspace{-1em}
\begin{tabular}{|c|}
\hline
$t_1$.S \\
\hline
185 \\
785 \\
\hline
\end{tabular}
\times
\begin{tabular}{|c|}
\hline
$t_1$.N \\
\hline
Smith \\
\hline
\end{tabular}
\times
\begin{tabular}{|c|}
\hline
$t_1$.M \\
\hline
1 \\
2 \\
\hline
\end{tabular}
\times
\begin{tabular}{|c|}
\hline
$t_2$.S \\
\hline
185 \\
186 \\
\hline
\end{tabular}
\times
\begin{tabular}{|c|}
\hline
$t_2$.N \\
\hline
Brown \\
\hline
\end{tabular}
\times
\begin{tabular}{|c|}
\hline
$t_2$.M \\
\hline
1 \\
2 \\
3 \\
4 \\
\hline
\end{tabular}
\]
}
% \vspace{-3mm}
%\caption{World-set decomposition for the initial or-set relation.}
%\label{fig:initialdec}
%\end{figure}

\begin{example}
\label{ex:1}
 \em
In the same way we can represent the result of data cleaning with the
uniqueness constraint for the social security numbers
as the product of Figure~\ref{fig:dec}.

% \vspace{-4mm}
\begin{figure}[h!]
	{\footnotesize
		\[
			%\hspace{-1mm}
			\begin{tabular}{|c@{\extracolsep{1mm}}c|}
				\hline
				$t_1$.S & $t_2$.S \\
				\hline
				185 & 186 \\
				785 & 185 \\
				785 & 186 \\
				\hline
			\end{tabular}
			\times
			\begin{tabular}{|c|}
				\hline
				$t_1$.N \\
				\hline
				Smith \\
				\hline
			\end{tabular}
			\times
			\begin{tabular}{|c|}
				\hline
				$t_1$.M \\
				\hline
				1 \\
				2 \\
				\hline
			\end{tabular}
			\times
			\begin{tabular}{|c|}
				\hline
				$t_2$.N \\
				\hline
				Brown \\
				\hline
			\end{tabular}
			\times
			\begin{tabular}{|c|}
				\hline
				$t_2$.M \\
				\hline
				1 \\
				2 \\
				3 \\
				4 \\
				\hline
			\end{tabular}
		\]
	}
	% \vspace{-6mm}
	\caption{WSD of the relation in Figure~\ref{fig:bigtable}.}
	% \vspace{-4mm}
	\label{fig:dec}
\end{figure}

One can observe that the result of this product is exactly the world-set
relation in Figure \ref{fig:bigtable}.
The presented decomposition is based on the {\em independence}\/
between sets of fields, subsequently called {\em components}\/.
Only fields that depend on
each other, for example $t_1.S$ and $t_2.S$, belong to the same
component. Since $\{ t_1.S, t_2.S \}$ and $\{ t_1.M \}$ are independent,
they are put into different components.
\punto
\end{example}

Often, one can quantify the certainty of a combination of possible values using probabilities.
For example, an automatic extraction tool that extracts structured data from text can produce a ranked list of possible extractions, each associated with a probability of being the correct one~\cite{gs06vldb}.

WSDs can elegantly handle such scenarios by simply adding a new column Pr to each component relation, which contains the probability for the corresponding combination of values.

\begin{figure}[h!]{
	\footnotesize
	\begin{center}
		%\hspace{-1mm}
		\begin{tabular}{|@{~}c@{~}c@{~}|@{~}c@{~}|}
			\hline
			$t_1$.S & $t_2$.S & Pr \\
			\hline
			185 & 186 & 0.2\\
			785 & 185 & 0.4 \\
			785 & 186 & 0.4 \\
			\hline
		\end{tabular}
		$\times$
		\begin{tabular}{|@{~}c@{~}|@{~}c@{~}|}
			\hline
			$t_1$.N & Pr \\
			\hline
			Smith & 1\\
			\hline
		\end{tabular}
		$\times$
		\begin{tabular}{|@{~}c@{~}|@{~}c@{~}|}
			\hline
			$t_1$.M & Pr \\
			\hline
			1 & 0.7 \\
			2 & 0.3 \\
			\hline
		\end{tabular}
		$\times$
		\begin{tabular}{|@{~}c@{~}|@{~}c@{~}|}
			\hline
			$t_2$.N & Pr \\
			\hline
			Brown & 1 \\
			\hline
		\end{tabular}
		$\times$
		\begin{tabular}{|@{~}c@{~}|@{~}c@{~}|}
			\hline
			$t_2$.M & Pr \\
			\hline
			1 & 0.25 \\
			2 & 0.25 \\
			3 & 0.25 \\
			4 & 0.25 \\
			\hline
		\end{tabular}
	\end{center}
	}
	% \vspace{-6mm}
	\caption{Probabilistic version of the WSD of Figure~\ref{fig:dec}.}
	% \vspace{-4mm}
	\label{fig:prob-wsd}
\end{figure}

\begin{example}
	\em
	\label{ex:prob-wsd}
	Figure~\ref{fig:prob-wsd} shows a probabilistic version of the WSD of Figure~\ref{fig:dec}. The probabilities in the last component imply that the possible values for the marital status of $t_2$ are equally likely, whereas $t_1$ is more likely to be single than married. The probabilities for the name values for $t_1$ and $t_2$ equal one, as this information is certain. 
	\punto
\end{example}

Given a probabilistic WSD $\{{\cal C}_1,\ldots,{\cal C}_m\}$, we obtain a possible world by choosing one tuple $w_i$ out of each component relation ${\cal C}_i$. The probability of this world is then computed as $\prod\limits_{i}w_i.Pr$. For example, in Figure~\ref{fig:prob-wsd} choosing the first, the second and the third tuple from the first, the third and the fifth component, respectively, results in the world
\begin{center}
	{\footnotesize
		\begin{tabular}{c|ccc}
			R & SSN & Name & MS \\
			\hline
			$t_1$ & 185 & Smith & 2 \\
			$t_2$ & 186 & Brown & 2
		\end{tabular}
	}
\end{center}
The world's probability can be computed as $0.2\cdot 0.3\cdot 0.25 = 0.015$.

In practice, it is often the case that fields or even tuples carry the
same values in all worlds. For instance, in the census data scenario
discussed above, we assumed that only one field in 10000 has several
possible values.  Such a world-set decomposes into a WSD in which most
fields are in component relations that have precisely one tuple.

We will also consider a refinement of WSDs, {\em WSDTs}\/,
which store information that is the same in
all possible worlds once and for all in so-called {\em template relations}\/.

\begin{example}
\label{ex:wsdt}
 \em
The world-set of the previous examples can be represented by the WSDT
of Figure~\ref{fig:wsdt}. \punto
\end{example}

% \vspace{-8mm}

\begin{figure}[h!]
	\centering
	{\footnotesize
	\begin{tabular}{c|c@{\extracolsep{1mm}}c@{\extracolsep{1mm}}c}
	Template & S & N & M \\
	\hline
	$t_1$ & ? & Smith & ? \\
	$t_2$ & ? & Brown & ? \\
	\end{tabular}
	\\
		\begin{tabular}{|@{~}c@{~}c@{~}|@{~}c@{~}|}
			\hline
			$t_1$.S & $t_2$.S & Pr \\
			\hline
			185 & 186 & 0.2\\
			785 & 185 & 0.4 \\
			785 & 186 & 0.4 \\
			\hline
		\end{tabular}
		$\times$
		\begin{tabular}{|@{~}c@{~}|@{~}c@{~}|}
			\hline
			$t_1$.M & Pr \\
			\hline
			1 & 0.7 \\
			2 & 0.3 \\
			\hline
		\end{tabular}
		$\times$
		\begin{tabular}{|@{~}c@{~}|@{~}c@{~}|}
			\hline
			$t_2$.M & Pr \\
			\hline
			1 & 0.25 \\
			2 & 0.25 \\
			3 & 0.25 \\
			4 & 0.25 \\
			\hline
		\end{tabular}
	}
	
	% \vspace{-4mm}
	\caption{Probabilistic WSD with a template relation.}
	% \vspace{-2mm}
	\label{fig:wsdt}
\end{figure}

% Using WSDs and WSDTs we can represent sets of worlds that cannot be expressed
% with or-set relations; at the same time the decomposition is just slightly
% larger than the original or-set relation for those cases where world-sets are
% actually expressible as or-set relations and can be compared.

% \begin{figure}[ph!]
%   \begin{center}
% \begin{small}
% \begin{tabular}{c|c@{\extracolsep{1mm}}c@{\extracolsep{1mm}}cl}
% c-table & S & N & M & \\
% \hline
%       &   &       &   & $(x = 185 \land z = 186) \lor$ \\
%       &   &       &   & $(x = 785 \land z = 185) \lor$ \\
%       &   &       &   & $(x = 785 \land z = 186)$ \\
% \hline
%  & $x$ & Smith & $y$ & $y = 1 \lor y = 2$\\
%  & $z$ & Brown & $w$ & $w = 1 \lor w = 2 \lor w = 3 \lor w = 4$\\
% \end{tabular}
% \end{small}
% \end{center}
% \vspace{-4mm}

%\begin{figure}[h!]
%\vspace{-4mm}

%\caption{A c-table T with global condition $\Phi$ encoding the WSDT of Figure~\ref{fig:wsdt}.}
%\vspace{-1em}
%\label{fig:c-table}
%\end{figure}

WSDTs combine the advantages of WSDs and c-tables. In fact, WSDTs can
be naturally viewed as c-tables whose formulas have been put into a
{\em normal form}\/ represented by the component relations, and
null values `?' in the template relations
represent fields on which the worlds disagree. Indeed, each
tuple in the product of the component relations is a possible value
assignment for the variables in the template relation. The following
c-table with global condition $\Phi$ is equivalent to the WSDT in Figure~\ref{fig:wsdt} (modulo the probabilistic weights):
% Fields at
% which the possible worlds disagree are assigned variables, and the
% global condition encodes the possible variable assignments.

% \vspace{-3mm}
{\footnotesize
  \begin{center}
  \begin{tabular}{c|c@{\extracolsep{1mm}}c@{\extracolsep{1mm}}cl}
T & S & N & M & \\
\hline
 & $x$ & Smith & $y$ & \\
 & $z$ & Brown & $w$ & \\
\end{tabular}
\begin{eqnarray}
\nonumber
\Phi & = & ((x = 185 \land z = 186) \lor (x = 785 \land z = 185) \lor \\\nonumber
                        && (x = 785 \land z = 186)) \land (y = 1 \lor y = 2)\land \\\nonumber
                        && (w = 1 \lor w = 2 \lor w = 3 \lor w = 4)
\end{eqnarray}

\end{center}
}

%% In the c-table

%%  the placeholders are replaced by variables and the possible
%% values for the variables are encoded with the conditions in the last column of
%% the table. The condition in the top row refers to more than one tuple; it
%% corresponds to the first component in our WSDT. The remaining two conditions
%% are local and are equivalent to the information given in the second and the
%% third component of the WSDT, respectively.

%\subsection*{Contributions}

The technical contributions of this article are as follows.
\begin{itemize}
%\addtolength{\itemsep}{-1.5ex} 
\item
We formally introduce WSDs and WSDTs and study some of their properties.
Our notion is a refinement of the one presented above and allows to represent
worlds over multi-relation schemas which contain relations with varying
numbers of tuples.
WSD(T)s can represent any finite set of possible worlds over
relational databases and are therefore a strong representation system for
{\em any relational query language}\/.

\item
A practical problem with WSDs and WSDTs is that
a DBMS that manages such representations has to support relations of arbitrary
arity: the schemata of the component relations of a decomposition
depend on the data.
Unfortunately, DBMS (e.g.\ PostgreSQL) in practice often do not
support relations beyond a fixed arity.

For that reason we present refinements of the notion of WSDs,
the {\em uniform WSDs (UWSDs)}\/, and their extension by template relations,
the {\em UWSDTs}\/, and study
their properties as representation systems.

\item We show how to process relational algebra queries over
  world-sets represented by UWSDTs.  For illustration purposes, we
  discuss query evaluation in the context of the much more
  graphic WSDs.
  
  We also develop a number of optimizations and techniques for
  normalizing the data representations obtained by queries to support
  scalable query processing even on very large world-sets.

\item
We describe a prototype implementation built on top of the PostgreSQL
RDBMS. Our system is called MayBMS and supports the management of incomplete information using UWSDTs.

\item We report on our experimental evaluation of UWSDTs as a representation
  system for large finite sets of possible worlds. Our experiments show that
  UWSDTs allow highly scalable techniques for managing incomplete information.
  We found that the size of UWSDTs obtained as query answers or data cleaning
  results remains close to that of a single world.  Furthermore, the processing time
  for queries on UWSDTs is also comparable to processing just a single world
  and thus a classical relational database.
  
\item For our experiments, we develop data cleaning techniques  in
  the context of UWSDTs. To clean data of inconsistent worlds
  we chase a set of equality-generating dependencies on UWSDTs,
  which we {brief-ly} describe.

%
%We
%focus on two kinds of dependencies, functional dependencies and a class of
%(in)equality-generating dependencies, and adapt the {\em Chase procedure}\/
%(cf.\ \cite{MMS79,ABU79,Gra1984}) for 
%incomplete information to the framework of UWSDTs.
\end{itemize}

WSDs are designed to cope with large sets of worlds, which exhibit local
dependencies and large commonalities. Note that this data pattern can be found
in many applications. Besides the census scenario, Section~\ref{sec:applications}
describes two further applications: managing
inconsistent databases using minimal repairs~\cite{ABC1999,BFFR2005} and medicine data.

A fundamental assumption of this work is that one wants to manage {\em finite
  sets of possible worlds}\/. This is justified by previous work on
representation systems starting with Imielinski and Lipski \cite{IL1984}, by
recent work ~\cite{dalvi04efficient,miller06clean,trio}, and by current
application requirements.
Our approach can deal with databases with unresolved uncertainties.
Such databases are still valuable. It should be possible to do data transformations that 
preserve as much information as possible, thus necessarily mapping between sets of possible worlds.
In this sense, WSDs represent a {\em compositional framework}\/ for querying and data cleaning.
%For example, data cleaning is often an incremental process that requires to store
%large intermediate results, world-sets, in databases.
%
A different approach is followed in, e.g.,~\cite{ABC1999,cali03}, where the
focus is on finding {\em certain answers}\/ of queries on incomplete and
inconsistent databases.

%Regarding data cleaning, we demonstrate that it is
%well feasible on WSDs. In our experiments, data cleaning provides
%us some interesting
%world-sets to run queries on. Further work will be required to close
%the gap between our simple dependency-theoretic framework and the state-of-the
%art in data cleaning \cite{RD2000, GFSS2000, RH2001, CGGM2003, BFFR2005},
%but we hope
%that the present paper renders it plausible to the reader that the framework
%of WSDs is relevant for data cleaning as well.

% This is self-defeating and I don't understand the point of it. It's also
% a mix of unrelated issues.
%
%Differently from c-tables, WSDs cannot represent infinite world-sets.
%A second assumption of this work is that {\em finite}\/ world-sets are
%relevant.  Furthermore, as discussed above, (U)WSDTs can be seen as a
%normal form for the c-tables representing finite world-sets.  (U)WSDTs
%support efficient representation and processing, with the promise of
%yet more expressive while still scalable representation systems to
%follow in the future.

\smallskip

\noindent{\underline{\bf Related Work.}}
%
%Recent years have witnessed important contributions towards scalable methods
%for managing incompleteness in relational databases. This section compares
%expressiveness and processing aspects of the contributions
%\cite{dalvi04efficient,miller06clean,trio} with the WSDs approach.
%
The probabilistic databases of \cite{dalvi04efficient,udbms05} and the dirty
relations of \cite{miller06clean} are examples of practical
representation systems that are not strong for relational algebra. 
As query answers in general cannot be represented as a set of possible
worlds in the same formalism, query evaluation is focused on computing
the certain answers to a query, or the probability of a tuple being in
the result.
Such formalisms close the possible worlds semantics using clean
answers~\cite{miller06clean} and probabilistic-ranked
retrieval~\cite{dalvi04efficient}.
%Thus, differently from WSDs, they do
%not consider the challenge of scalable and strong representation systems.
As we will see in this article, our approach subsumes the aforementioned two and is strictly more expressive than them.

%In contrast to or-set relations, Miller's representation system contain
%alternatives for whole tuples rather than for single fields of a tuple only.
%Its salient weakness is the inability (1) to efficiently represent
%alternatives for single fields (like or-set relations do), and (2) to
%represent alternatives for sets of fields across several tuples (like WSDs
%do). To see the former point, consider an or-set relation with one tuple,
%which has $m$ or-sets, each with $n$ values. The number of alternatives of the
%same tuple is then $n^m$, i.e, the number of possible worlds.

%The clean answer of a query is defined as the set of tuples that are possible
%answers of that query, together with their probabilities. Miller et al.\@
%further investigate the query evaluation for the restricted class of queries
%with acyclic join graphs.

%Tuple-set independence can be easily modeled in WSDs by having a component for
%each set of alternatives (which become component's tuples).  

%Note that the or-set, tuple, and tuple-set independence assumptions discussed
%above drastically simplify the query evaluation on WSDs, because each tuple of
%a database relation becomes now a tuple in a component. Component merging, the
%most expensive operation on WSDs, is thus only needed when joining independent
%relations.

In parallel to our approach, \cite{widom06working,trio}
propose ULDBs that combine uncertainty and a low-level form of
lineage to model any finite world-set.  Like the dirty relations of
\cite{miller06clean}, ULDBs represent a set of independent tuples with
alternatives. Lineage is then used to represent dependencies among
alternatives of different tuples and thus is essential for the
expressive power of the formalism.

As both ULDBs and WSDs can model any finite world-set, they inherently
share some similarities, yet differ in important aspects. WSDs support
efficient algorithms for finding a minimal data representation based
on relational factorization. Differently from ULDBs,\\
WSDs allow
representing uncertainty at the level of tuple fields, not only of tuples. This
causes, for instance, or-set relations to have linear representations
as WSDs, but (in general) exponential representations as ULDBs.  As
reported in \cite{trio}, resolving tuple dependencies, i.e., tracking
which alternatives of different tuples belong to the same world, often
requires to compute expensive lineage closure.  Additionally, query
operations on ULDBs can produce inconsistencies and anomalies, such as
erroneous dependencies and inexistent tuples. In contrast, WSDs share
neither of these pitfalls.  As no implementation of ULDBs was available at the time of writing this document, no experimental comparison of ULDBs and WSDs could be established. 

\section{Preliminaries}
\label{sec:preliminaries}

We use the named perspective of the relational model with the operations
selection $\sigma$, projection $\pi$, product $\times$, union $\cup$,
difference $-$, and attribute renaming $\delta$ (cf.\ e.g.\ \cite{AHV95}).  A
{\em relational schema}\/ is a tuple $\Sigma = (R_1[U_1], \ldots, R_k[U_k])$, where each $R_i$ is a relation name and $U_i$ is a set of attribute names.  Let ${\bf
  D}$ be a set of domain elements.  A {\em relation}\/ over schema $R[A_1,
\dots, A_k]$ is a set of tuples $(A_1: a_1, \dots, A_k: a_k)$ where $a_1,
\dots, a_k \in {\bf D}$.  A {\em relational database}\/ ${\cal A}$ over schema
$\Sigma$ is a set of relations $R^{\cal A}$, one for each relation schema
$R[U]$ from $\Sigma$.  Sometimes, when no confusion of database may occur, we
will use $R$ rather than $R^{\cal A}$ to denote one particular relation over
schema $R[U]$.  By the size of a relation $R$, denoted $|R|$, we refer to the
number of tuples in $R$. For a relation $R$ over schema $R[U]$, let
$\mathcal{S}(R)$ denote the set $U$ of its attributes and let $ar(R)$ denote
the arity of $R$.

A {\em product $m$-decomposition} of a relation $R$ is a set of
non-nullary relations $\{C_1,\ldots,C_m\}$ such that
$C_1\times\cdots\times C_m = R$. The relations $C_1,\ldots,C_m$ are
called {\em components}. A product $m$-decomposition of $R$ is {\em
  maximal} if there is no product $n$-decomposition of $R$ with $n>m$.

A set of {\em possible worlds}\/ (or {\em world-set}) over schema
$\Sigma$ is a set of databases over schema $\Sigma$. Let ${\bf W}$ be
a set of structures, $rep$ be a function that maps from ${\bf W}$ to world-sets of
the same schema. Then $({\bf W}, rep)$ is a {\em strong representation system}\/
for a query language if, for each query $Q$ of that language and each
${\cal W} \in {\bf W}$ such that $Q$ is applicable to the worlds in
$rep({\cal W})$, there is a structure ${\cal W}' \in {\bf W}$ such
that $ rep({\cal W}') = \{ Q({\cal A}) \mid {\cal A} \in rep({\cal W})
\}$.  Obviously,

\begin{lemma}
\label{lem:finite_strong}
If $rep$ is a function from a set of structures ${\bf W}$ to the set of all finite world-sets,
then $({\bf W}, rep)$ is a strong representation system for any relational query language.
\end{lemma}

%% For the remainder of the paper we consider that
%% the arity of all our database relations is at least one and the projection
%% operation does not project to the empty set of attributes.

%%% Local Variables: 
%%% mode: latex
%%% TeX-master: "main"
%%% End: 

\section{Probabilistic World-Set Decompositions}
\label{sec:wsd}

%\noindent\textbf{The Basic Notion.} 

In order to use classical database techniques for storing and querying
incomplete data, we develop a scheme for representing a world-set ${\bf A}$ by
a single relational database.

Let ${\bf A}$ be a finite world-set over schema $\Sigma = $
$(R_1, \ldots, R_k)$. 
For each $R$ in $\Sigma$, let $|R|_{\max} = \max \{ |R^{\cal A}| : {\cal
  A} \in {\bf A} \}$ denote the maximum cardinality of relation $R$ in any
world of ${\bf A}$.  Given a world ${\cal A}$ with $R^{\cal A} = \{ t_1,
\dots, t_{|R^{\cal A}|} \}$, let $t_{R^{\cal A}}$ be the tuple obtained as the
concatenation (denoted $\circ$) of the tuples of $R^{\cal A}$ in an arbitrary
order padded with a special tuple $t_{\bot}=\underbrace{(\bot, \dots, \bot)}_{ar(R)}$ up to arity
$|R|_{\max}$:
\[
t_{R^{\cal A}} :=
t_1 \circ \dots \circ t_{|R^{\cal A}|} \circ
(\underbrace{t_{\bot}, \dots\dots\dots\dots, t_{\bot}}_{|R|_{\max} -
  |R^{\cal A}|})
\]
Then tuple $t_{\cal A} := t_{R_1^{\cal A}} \circ \dots \circ t_{R_k^{\cal A}}$
encodes all the information in world ${\cal A}$. The ``dummy'' tuples with
$\bot$-values are only used to ensure that the relation $R$ has the same
number of tuples in all worlds in \textbf{A}. We extend this interpretation and generally define as $t_{\bot}$ any tuple
that has at least one symbol $\bot$, i.e., $(A_1 : a_1, . . . ,A_n : a_n)$, where at least one $a_i$
is $\bot$, is a $t_{\bot}$ tuple. This allows for several different inlinings of the same world-set.

By a {\em world-set relation}\/ of a world-set ${\bf A}$, we denote the
relation $\{ t_{\cal A} \mid {\cal A} \in {\bf A} \}$.  This world-set
relation has schema $ \{ R.t_i.A_j \mid R[U] \in \Sigma, 1 \le i \le
|R|_{\max}, A_j \in U \}$. Note that in defining this schema we use $t_i$ to
denote the position (or identifier) of tuple $t_i$ in $t_{R^{\cal A}}$ and
not its value.

Given the above definition that turned every world in a tuple of a
world-set relation, computing the initial world-set is an easy exercise. In
order to have every world-set relation define a world-set, let a tuple
extracted from some $t_{R^{\cal A}}$ be in $R^{\cal A}$ iff it does not
contain any occurrence of the special symbol $\bot$.  That is, we map
$t_{R^{\cal A}} = (a_1, \dots, a_{ar(R) \cdot |R|_{\max}})$ to $R^{\cal A}$ as
\begin{multline*}
t_{R^{\cal A}} \mapsto
\{
(a_{ar(R) \cdot k + 1}, \dots, a_{ar(R) \cdot (k+1)}) \mid
   0 \le k < |R|_{\max}, \\
   a_{ar(R) \cdot k + 1} \neq \bot, \dots, a_{ar(R) \cdot (k+1)}  \neq \bot
\}.
\end{multline*}
%% (So one can think of $\bot$ as a deletion marker for
%% tuples.)

Observe that although world-set relations are not unique as we have
left open the ordering in which the tuples of a given world are
concatenated, all world-set relations of a world-set ${\bf A}$ are
equally good for our purposes because they can be mapped invariantly
back to ${\bf A}$. Note that for each world-set relation a maximal
decomposition exists, is unique, and can be efficiently
computed~\cite{ako06worlds}.

\begin{definition}
  \em Let ${\bf A}$ be a world-set and $W$ a world-set relation representing
  ${\bf A}$.  Then a {\em world-set $m$-decomposition ($m$-WSD)}\/ of ${\bf
    A}$ is a product $m$-decomposition of $W$.
\end{definition}

We will refer to each of the $m$ elements of a world-set $m$-decomposition as {\em components},
and to the component tuples as {\em local worlds}.
Somewhat simplified examples of world-set relations and WSDs over a single
relation $R$ (thus ``$R$'' was omitted from the attribute names of the
world-set relations) were given in Section~\ref{sec:introduction}. Further
examples can be found in Section~\ref{sec:queries}.
It should be emphasized that with WSDs we can also represent multiple
relational schemata and even components with fields from different
relations.
% Approaches like lossless join decompositions~\cite{AHV95}
% differ from WSDs in that they are not defined intensionally using
% dependencies.

It immediately follows from our definitions that

\begin{proposition}
Any finite set of possible worlds can be represented as a world-set
relation and as a $1$-WSD.
\end{proposition}

\begin{corollary}[Lemma~\ref{lem:finite_strong}]
WSDs are a strong re\-pre\-sen\-tation system for any relational
query language.
\end{corollary}

As pointed out in Section~\ref{sec:introduction}, this is not true for or-set
relations. For the relatively small class of world-sets that can be
represented as or-set relations, the size of our representation system is
linear in the size of the or-set relations. As seen in the examples, our
representation is {\em much more space-efficient than world-set relations}.

%   Decompositions of extensionally given relations are less natural in
%   the classical context because such decompositions may break on
%   updates.  

\nop{ % removed for ICDE
\noindent {\bf Proof}.
Existence is clear because a world-set relation is also a 1-WSD.
Uniqueness is shown by contradiction: Given relation $R$ of schema $R[U]$,
assume that there are two different maximal $m$-de\-com\-po\-si\-tions
$\{ U_1, \dots, U_m \}$ and $\{ V_1, \dots, V_m \}$ of $R$.
Since the two decompositions are different,
there are two sets $U_i, V_j$ such that $U_i \neq V_j$ and $U_i \cap V_j \neq
\emptyset$.
But then, as of course
$R = \pi_{U - V_j}(R) \times \pi_{V_j}(R)$,
we have
$
\pi_{U_i}(R) =
\pi_{U_i}\big(\pi_{U - V_j}(R) \times \pi_{V_j}(R)\big)
= \pi_{U_i-V_j}(R) \times \pi_{U_i \cap V_j}(R).
$
It follows that
$
\{ U_1, \dots, U_{i-1}, U_i - V_j, U_i \cap V_j, U_{i+1}, \dots, U_m \}
$
is an $(m+1)$-decomposition of $R$, and $m$-decompositions cannot be
maximal. Contradiction.
\punto
}
%\vspace*{1em}

%%%%%%%%%%%%%%%%%%%%%%%%%%%%%%%%%%%%%%%%%%%%%%%%%%%%%%%%%%%%%%%%%%%%%
%%%%%%%%%%%%%%%%%%%%%%%%%%%%%%%%%%%%%%%%%%%%%%%%%%%%%%%%%%%%%%%%%%%%%
\bigskip

\noindent\textbf{Modeling Probabilistic Information.}
We can quantify the uncertainty of the data by means of probabilities using a natural extension of WSDs. A {\em probabilistic world-set m-decomposition} (probabilistic m-WSD) is an m-WSD $\{{\cal C}_1,\ldots,{\cal C}_m\}$, where each component relation ${\cal C}$ has a special attribute $Pr$ in its schema defining the probability for the local worlds, that is, for each combination of values defined by the component. We require that the probabilities in a component sum up to one, i.e.\ $\sum\limits_{t_C\in{\cal C}}t_C.Pr=1$. 

Probabilistic WSDs generalize the probabilistic tuple-independent model of \cite{dalvi04efficient}, as we show next.
Figure~\ref{fig:dalvi-db}~(a) is an example taken from \cite{dalvi04efficient}. It shows a probabilistic database with two relations $S$ and $T$. Each tuple is assigned a
confidence value, which represents the probability of the tuple being
in the database, and the tuples are assumed independent.
A possible world is obtained by choosing a subset of the tuples in the probabilistic database, and its
probability is computed by multiplying the probabilities for selecting a tuple or not, depending on whether
that tuple is in the world. The set of possible worlds for $D$ is given in
Figure~\ref{fig:dalvi-db}~(b). For example, the probability of the world $D_3$ can be computed as $(1-0.2)\cdot0.5\cdot0.6=0.06$.

\begin{figure}[h]
  \centering
  \begin{tabular}{cc}
	  \begin{tabular}{c}
		  \begin{tabular}{c|cc|c}
		    S & A & B & Pr \\\hline
		    $s_1$  & m & 1 & 0.8\\
		    $s_2$  & n & 1 & 0.5\\
		  \end{tabular}\\\\
		  \begin{tabular}{c|cc|c}
		    T & C & D & Pr \\\hline
		    $t_1$  & 1 & p & 0.6\\
		  \end{tabular}\\\\
		  (a)
	  \end{tabular}
	  &
	  \begin{tabular}{c}
		  \begin{tabular}{|l|c}
		    world & Pr \\\hline
		    $D_1 = \{s_1,s_2,t_1\}$ & 0.24\\
		    $D_2 = \{s_1,t_1\}$     & 0.24\\
		    $D_3 = \{s_2,t_1\}$     & 0.06\\
		    $D_4 = \{t_1\}$         & 0.06\\
		    $D_5 = \{s_1,s_2\}$     & 0.16\\
		    $D_6 = \{s_1\}$         & 0.16\\
		    $D_7 = \{s_2\}$         & 0.04\\
		    $D_8 = \emptyset$       & 0.04\\    
		  \end{tabular}\\\\
		  (b)
	  \end{tabular}
  \end{tabular}
  \caption{A probabilistic database for relations S and T (a), and the represented set of possible worlds (b).}
  \label{fig:dalvi-db}
\end{figure}

We obtain a probabilistic WSD in the following way. Each tuple $t$ with
confidence $c$ in a probabilistic database induces a WSD component
representing two local worlds: the local world with tuple $t$ and
probability $c$, and the empty world with probability  $1-c$.
Figure~\ref{fig:dalvi-wsd} gives the WSD encoding of the
probabilistic database of Figure~\ref{fig:dalvi-db}. Of course, in probabilistic WSDs we can assign probabilities not only to individual tuples, but also to combinations of values for fields of different tuples or relations.

\begin{figure}[thbp]
	\centering
  \begin{small}
	\parbox{\textwidth}{
    \begin{tabular}{|@{~}c@{~}|@{~}c@{~}c@{~}|@{~}l@{~}|}
      \hline
      $C_1$ & $s_1$.A & $s_1$.B & Pr \\\hline
      1 & m      & 1      & 0.8\\
      2 & $\bot$ & $\bot$ & 0.2\\\hline
    \end{tabular}
    $\times$
    \begin{tabular}{|@{~}c@{~}|@{~}c@{~}c@{~}|@{~}l@{~}|}
      \hline
      $C_2$ & $s_2$.A & $s_2$.B & Pr \\\hline
      1 & n      & 1      & 0.5\\
      2 & $\bot$ & $\bot$ & 0.5\\\hline
    \end{tabular}
    $\times$
    \begin{tabular}{|@{~}c@{~}|@{~}c@{~}c@{~}|@{~}l@{~}|}
      \hline
      $C_3$ & $t_1$.C & $t_1$.D & Pr \\\hline
      1 & 1      & p      & 0.6\\
      2 & $\bot$ & $\bot$ & 0.4\\\hline
    \end{tabular}
    }
\end{small}
\caption{WSD equivalent to the probabilistic database in Figure~\ref{fig:dalvi-db}~(a).}
\label{fig:dalvi-wsd}
\end{figure}

%%%%%%%%%%%%%%%%%%%%%%%%%%%%%%%%%%%%%%%%%%%%%%%%%%%%%%%%%%%%%%%%%%%%%
%%%%%%%%%%%%%%%%%%%%%%%%%%%%%%%%%%%%%%%%%%%%%%%%%%%%%%%%%%%%%%%%%%%%%

\noindent\textbf{Adding Template Relations.} We now present our refinement
of WSDs with so-called {\em template relations}. A template stores information
that is the same in all possible worlds and contains special values `$?$'
$\notin$ \textbf{D} in fields at which different worlds disagree.

%% We will assume that tuples $t$ have unique ids $\hat{t}$. Let $\Sigma = \{
%% R_1, \dots, R_k \}$ be a schema and ${\bf A}$ a finite set of possible worlds
%% over $\Sigma$. Then, the database $(R_1^0, \dots, R_k^0, \{C_1, \dots, C_m\})$
%% is called an {\em $m$-WSD with template relations ($m$-WSDT)}\/ of ${\bf A}$
%% iff there is a WSD $\{C_1, \dots, C_m, D_1, \dots, D_n\}$ of ${\bf A}$ such
%% that $|D_i| = 1$ for all $i$ and if relation $D_i$ has attribute
%% $R_j.\hat{t}.A$ and value $v$ in its unique $R_j.\hat{t}.A$-field, then the
%% template relation $R_j^0$ has a tuple with id $\hat{t}$ whose $A$-field has
%% value $v$.

Let $\Sigma = (R_1, \dots, R_k)$ be a schema and ${\bf A}$ a finite set of
possible worlds over $\Sigma$. Then, the database $(R_1^0, \dots, R_k^0,
\{C_1, \dots, C_m\})$ is called an {\em $m$-WSD with template relations
  ($m$-WSDT)}\/ of ${\bf A}$ iff there is a WSD $\{C_1, \dots, C_m, D_1,
\dots, D_n\}$ of ${\bf A}$ such that $|D_i| = 1$ for all $i$ and if relation
$D_i$ has attribute $R_j.t.A$ and value $v$ in its unique $R_j.t.A$-field,
then the template relation $R_j^0$ has a tuple with identifier $t$ whose
$A$-field has value $v$.

Of course WSDTs again can represent any finite world-set and are thus a strong
representation system for any relational query language. 
%% Note that in contrast to
%% WSDs, WSDTs admit a unique maximal decomposition only if the template relation
%% $R^0$ is fixed.
Example~\ref{ex:wsdt} shows a WSDT for the running example of the
introduction.

%\vspace*{1em}

%%%%%%%%%%%%%%%%%%%%%%%%%%%%%%%%%%%%%%%%%%%%%%%%%%%%%%%%%%%%%%%%%%%%%
%%%%%%%%%%%%%%%%%%%%%%%%%%%%%%%%%%%%%%%%%%%%%%%%%%%%%%%%%%%%%%%%%%%%%

\noindent\textbf{Uniform World-Set Decompositions.} In practice database systems
often do not support relations of arbitrary arity (e.g., WSD components).  For
that reason we introduce next a modified representation of WSDs called {\em
  uniform WSDs}.
Instead of having a
variable number of component relations, possibly with different arities, we
store all values in a single relation $C$ that has a fixed schema.
We use the fixed schema consisting of the three relation schemata
\[
	C[\mathit{FID},
	\mathit{LWID}, \mathit{VAL}], F[\mathit{FID}, \mathit{CID}], W[\mathit{CID}, \mathit{LWID}, \mathit{PR}]
\]
where $\mathit{FID}$ is a triple\footnote{That
  is, FID really takes three columns, but for readability we keep them
  together under a common name in this section.}  $(\mathit{Rel},
\mathit{TupleID}, \mathit{Attr})$ denoting the $\mathit{Attr}$-field of
tuple $TupleID$ in database relation $\mathit{Rel}$.

%\nop{ % removed for ICDE
In this representation we need a restricted flavor of world-ids called {\em
  local world-ids} (LWIDs). The local world-ids refer only to the possible
worlds within one component.  LWIDs avoid the drawbacks of ``global'' world
IDs for the individual worlds.  This is important, since the size of global
world IDs can exceed the size of the decomposition itself, thus making it
difficult or even impossible to represent the world-sets in a space-efficient
way. If any world-set over a given schema and a fixed active domain is
permitted, one can verify that global world-ids cannot be smaller than the
largest possible world over the schema and the active domain.
%}

Given a WSD $\{C_1, \dots, C_m\}$ with schemata $C_i[U_i]$, we populate the
corresponding UWSD as follows.
\begin{itemize}
%\addtolength{\itemsep}{-1ex} 
\item
$((R,t,A),s,v) \in C$ iff, for some (unique) $i$,
$R.t.A \in U_i$ and the field of column
$R.t.A$ in the tuple with id $s$ of $C_i$ has value $v$.
\item
$F := \{ ((R,t,A), C_i) \mid 
1 \le i \le m,\; R.t.A \in U_i \}$,

\item
  $(C_i, s, p) \in W$ iff there is a tuple with identifier
$s$ in $C_i$, whose probability is $p$.

\end{itemize}

Intuitively, the relation $C$ stores each value from a component together with its corresponding field identifier and the identifier of the component-tuple in the initial WSD (column $\mathit{LWID}$ of $C$). The relation $F$ contains the mapping between tuple fields and component identifiers, and $W$ keeps track of the worlds present for a given component.

%% Given a WSD $\{C_1, \dots, C_m\}$ with schemata $\hat{C_i}[U_i]$ (we now
%% distinguish between the relation and its name), we populate the corresponding
%% UWSD as follows.
%% \begin{itemize}
%% \addtolength{\itemsep}{-1ex} 
%% \item
%% $F := \{ ((R,\hat{t},A), \hat{C_i}) \mid 
%% 1 \le i \le m,\; R.\hat{t}.A \in U_i \}$,

%% \item
%% $(\hat{C_i}, \hat{s}) \in W$ iff there is a tuple with id
%% $\hat{s}$ in $C_i$.

%% \item
%% $((R,\hat{t}, A), \hat{s}, v) \in C$ iff, for some (unique) $i$,
%% $R.\hat{t}.A \in U_i$ and the field of column
%% $R.\hat{t}.A$ in the tuple with id $\hat{s}$ of $C_i$ has value $v$.
%% \end{itemize}

%\begin{remark}
%\em
In general, the VAL column in the component relation C must
store values for fields of different type. One possibility is
to store all values as strings and use casts when required.
Alternatively, one could have one component relation for each data type.
In both cases the schema remains fixed.

Finally, we add template relations to UWSDs in complete analogy with 
WSDTs, thus obtaining the UWSDTs.

\begin{figure}
	\begin{center}
		\hspace{-3mm}
		\begin{small}
			\begin{tabular}{c|c@{\extracolsep{1mm}}c@{\extracolsep{1mm}}c}
				$R^0$ & S & N & M \\
				\hline
				$t_1$ & ? & Smith & ? \\
				$t_2$ & ? & Brown & 3 \\
				\multicolumn{3}{c}{} \\
				C & FID & LWID & VAL \\
				\hline
				& $(R, t_1, S)$ & 1 & 185 \\
				& $(R, t_2, S)$ & 1 & 186 \\
				& $(R, t_1, S)$ & 2 & 785 \\
				& $(R, t_2, S)$ & 2 & 185 \\
				& $(R, t_1, S)$ & 3 & 785 \\
				& $(R, t_2, S)$ & 3 & 186 \\
				& $(R, t_1, M)$ & 1 & 1 \\
				& $(R, t_1, M)$ & 2 & 2 \\
			\end{tabular}
				%\hspace{-2mm}
			\begin{tabular}{c}
				\begin{tabular}{c|cc}
				F & FID & CID \\
				\hline
				& $(R, t_1, S)$ & $C_1$ \\
				& $(R, t_1, M)$ & $C_2$ \\
				& $(R, t_2, S)$ & $C_1$ 
				\end{tabular}\\\\
				\begin{tabular}{c|c@{\extracolsep{2mm}}c|@{\extracolsep{2mm}}c@{\extracolsep{2mm}}}
				W & CID & LWID & PR \\
				\hline
				& $C_1$ & 1 & 0.2 \\
				& $C_1$ & 2 & 0.4 \\
				& $C_1$ & 3 & 0.4 \\
				& $C_2$ & 1 & 0.7 \\
				& $C_2$ & 2 & 0.3 \\
				\end{tabular}
			\end{tabular}
		\end{small}
	\end{center}
	% \vspace{-2mm}
	\caption{A UWSDT corresponding to the WSDT of Figure~\ref{fig:wsdt}.}
	% \vspace{-4mm}
	\label{fig:uwsdt}
\end{figure}

\begin{example}
\em
 We modify the world-set represented in Figure \ref{fig:prob-wsd} such that
  the marital status in $t_2$ can only have the value 3. Figure~\ref{fig:uwsdt} is then the uniform version
of the WSDT of Figure~\ref{fig:prob-wsd}.
Here $R^0$ contains the values that are the same in all worlds. For each field
that can have more than one possible value, $R^0$ contains a special
placeholder, denoted by `$?$'. The possible values for the
placeholders are defined in the component table $C$. In practice,
we can expect that the majority of the
data fields can take only one value across all worlds, and can be stored in
the template relation.
\punto
\end{example}

% It is easy to verify that

\begin{proposition}
Any finite set of possible worlds can be represented as a $1$-UWSD and
as a $1$-UWSDT.
\end{proposition}

It follows again that UWSD(T)s are a strong representation system for
{\em any relational query language}\/.

%%% Local Variables: 
%%% mode: latex
%%% TeX-master: main
%%% TeX-master: "main"
%%% End: 

%-------------------------------

\section{Queries on World-set Decompositions}
\label{sec:queries}

In this section we study the query evaluation problem for WSDs. As
pointed out before, UWSDTs are a better representation system than
WSDs; nevertheless WSDs are simpler to explain and visualize and the
main issues regarding query evaluation are the same for both systems.

%For that reason we concentrate on query evaluation in the WSD context.

% Query evaluation for UWSDTs is then discussed in Section~\ref{sec:alg-scalable}.

The goal of this section is to provide, for each relational algebra query $Q$,
a query $\hat{Q}$ such that for a WSD ${\cal W}$,
\[
rep(\hat{Q}({\cal W})) = \{ Q({\cal A}) \mid {\cal A} \in rep({\cal W}) \}.
\]
Of course we want to evaluate queries directly on WSDs using $\hat{Q}$
rather than process the individual worlds using the original query $Q$.

The algorithms for processing relational algebra que-ries presented next are orthogonal to whether or not the WSD stores probabilities. According to our semantics, a query is conceptually evaluated in each world and extends the world with the result of the query in that world. A different class of queries are those that close the possible world semantics and compute {\em confidence} of tuples in the result of a query. This will be the subject of Section~\ref{sec:prob-wsd}.

When compared to traditional query evaluation, the evaluation of relational
queries on WSDs poses new challenges.  First, since decompositions in general
consist of several components, a query $\hat{Q}$ that maps from one WSD to
another must be expressed as a set of queries, each of which defines a
different component of the output WSD.  Second, as certain query operations
may cause new dependencies between components to develop, some components may
have to be merged (i.e., part of the decomposition undone using the product
operation $\times$). Third, the answer to a (sub)query $Q_0$ must be
represented within the same decomposition as the input relations; indeed, we
want to compute a decomposition of world set $\{({\cal A},Q_0({\cal A})) \mid
{\cal A} \in rep({\cal W}) \}$ in order to be able to resort to the input
relations as well as the result of $Q_0$ within each world.
Consider for example a query $\sigma_{A=1}(R) \cup \sigma_{B=2}(R)$.
If we first compute $\sigma_{A=1}(R)$, we must not replace $R$ by
$\sigma_{A=1}(R)$, otherwise $R$ will not be available for the computation of
$\sigma_{B=2}(R)$. On the other hand, if $\sigma_{A=1}(R)$ is stored in a
separate WSD, the connection between worlds of $R$ and the selection
$\sigma_{A=1}$ is lost and we can again not compute
$\sigma_{A=1}(R) \cup \sigma_{B=2}(R)$.

\begin{figure*}[t!]
\begin{small}
\framebox[\textwidth]{
\hspace{4mm}
\parbox{9cm}{
\begin{tabbing}
{\bf algorithm} select[$A \theta c$]
$\quad$ // compute $P := \sigma_{A \theta c} R$ \\
{\bf begin} \\
\hspace{2mm} \=
   $\textsf{copy}(R,P)$; \\
\> {\bf for each} $1 \le i \le |P|_{max}$ {\bf do} {\bf begin} \\
\> \hspace{2mm} \=
      {\bf let} $C$ be the component of $P.t_i.A$; \\
\> \> {\bf for each} $t_C \in C$ {\bf do}\\
\> \> \hspace{2mm} \=
            {\bf if not} ($t_C.(P.t_i.A) \,\theta\, c$) {\bf then} \\
\> \> \> \hspace{2mm} \=
                          $t_C.(P.t_i.A) := \bot$\\
\> \> $\textsf{propagate-$\bot$}(C)$; \\
\> {\bf end} \\
{\bf end}\\
\\
{\bf algorithm} select[$A \theta B$]
$\quad$ // compute $P := \sigma_{A \theta B} R$ \\
{\bf begin} \\
\hspace{2mm} \=
   $\textsf{copy}(R,P)$; \\
\> {\bf for each} $1 \le i \le |P|_{max}$ {\bf do} {\bf begin} \\
\> \hspace{2mm} \=
      {\bf let} $C$  be the component of $P.t_i.A$; \\
\> \> {\bf let} $C'$ be the component of $P.t_i.B$; \\
\> \> {\bf if} ($C \neq C'$) {\bf then} \\
\> \> \hspace{2mm} \=
       replace components $C$, $C'$ by $C := \textsf{compose}(C,C')$; \\
      % ${\cal C} := {\cal C} - \{ C, C' \} \cup \{ C \times C' \}$; \\
\> \> {\bf for each} $t_C \in C$ {\bf do}\\
\> \> \> {\bf if not} ($t_C.(P.t_i.A) \,\theta\, t_C.(P.t_i.B)$) {\bf then} \\
\> \> \> \hspace{2mm} \=
                                                $t_C.(P.t_i.A) := \bot$\\
%\> \> \> \> $t_C.(P.t_i.B) := \bot$\\
\> \> $\textsf{propagate-$\bot$}(C)$; \\
\> {\bf end} \\
{\bf end} \\
\\
{\bf algorithm} product
$\quad$ // compute $T := R \times S$ \\
{\bf begin} \\
\hspace{2mm} \=
   {\bf for each} $1 \le j \le |S|_{max}$ and $R.t_i.A\in\mathcal{S}(R)$  {\bf do} {\bf begin}~~ \\
\> \hspace{2mm} \=
      {\bf let} $C$ be the component of $R.t_i.A$; \\
\> \> $C := \textsf{ext}(C, R.t_i.A, T.t_{ij}.A)$; \\
\> {\bf end}; \\
\> {\bf for each} $1 \le i \le |R|_{max}$ and $S.t_j.A\in\mathcal{S}(S)$ {\bf do} {\bf begin} \\
\> \> {\bf let} $C'$ be the component of $S.t_j.A$; \\
\> \> $C' := \textsf{ext}(C', S.t_j.A, T.t_{ij}.A)$; \\
\> {\bf end} \\
{\bf end} \\
\\
{\bf algorithm} union
$\quad$ // compute $T := R \cup S$ \\
{\bf begin} \\
\hspace{2mm} \=
	{\bf for each} $1 \le i \le |R|_{max}$ and $A \in {\cal S}(R)$ {\bf do} {\bf begin} \\
\> \hspace{2mm} \=
 		{\bf let} $C$ be the component of $R.t_i.A$; \\
\> \> $C := \textsf{ext}(C, R.t_i.A, T.(R.t_i).A)$; \\
\> {\bf end}; \\
\> {\bf for each} $1 \le j \le |S|_{max}$ and $A \in {\cal S}(S)$ {\bf do} {\bf begin} \\
\> \> {\bf let} $C'$ be the component of $S.t_j.A$; \\
\> \> $C' := \textsf{ext}(C', S.t_j.A, T.(S.t_j).A)$; \\
\> {\bf end} \\
{\bf end}\\
\\
\end{tabbing}
}
\hfill
\parbox{9cm}{
\begin{tabbing}
{\bf algorithm} project[$U$]
$\quad$ // compute $P := \pi_U(R)$ \\
{\bf begin} \\
\hspace{2mm} \=
   $\textsf{copy}(R,P)$; \\
\> {\bf for each} $1 \le i \le |P|_{max}${\bf do} \\
\> \hspace{2mm} \=
      {\bf while} no fixpoint is reached {\bf do} {\bf begin}\\
\> \> \hspace{2mm} \=
         {\bf let} $C$ be the component of $P.t_i.A$, where $A\in U$; \\
\> \> \> {\bf let} $C'\not=C$ be the component of $P.t_i.B$, where \\
\> \> \> \hspace{2mm} \=
               $B\not\in U$ and ($\forall A'\in U:
               P.t_i.A'\notin\mathcal{S}(C')$) and \\
\> \> \> \>    ($\exists t_{C'}\in C': t_{C'}.B=\bot$);\\
\> \> \> replace components $C$, $C'$ by $C := \textsf{compose}(C,C')$;\\
\> \> \> $\textsf{propagate-$\bot$}(C)$; \\
\> \> \> project away $P.t_j.B$ from $C$ where $B\not\in U$ and $j\leq i$; \\
\> \> {\bf end} \\
\> {\bf for each} $1 \le i \le |P|_{max}$ and $B \notin U$ {\bf do} {\bf begin} \\
\> \> {\bf let} $C$ be the component of $P.t_i.B$; \\
\> \> project away $P.t_i.B$ from $C$; \\
\> {\bf end} \\
{\bf end}
\\\\
{\bf algorithm} rename
$\quad$ // compute $\delta_{A \rightarrow A'}(R)$ \\
{\bf begin} \\
\> {\bf for each} $1 \le i \le |R|_{max}$ {\bf do} {\bf begin} \\
\> \> {\bf let} $C$ be the component of $R.t_i.A$; \\
\> \> $C := \delta_{R.t_i.A \rightarrow R.t_i.A'}(C)$; \\
\> {\bf end}; \\
{\bf end}
\\
\\
{\bf algorithm} difference
$\quad$ // compute $P := R-S$ \\
{\bf begin} \\
\hspace{2mm} \=
   $\textsf{copy}(R,P)$; \\
\> {\bf for each} $1 \le i \le |P|_{max}${\bf do} \\
\> \> {\bf for each} $1 \le j \le |S|_{max}${\bf do} \\
\> \> \> {\bf let} $C_1,\ldots,C_k$ be the components for the fields of $P.t_i$ and $S.t_j$;\\
\> \> \> replace $C_1,\ldots,C_k$ by $C := \textsf{compose}(C_1,\ldots,C_k)$;\\
\>\>\> {\bf for each} $t_C \in C$ {\bf do} {\bf begin} \\
\>\>\>\> {\bf if} $t_C.(P.t_i.A)=t_C.(S.t_j.A)$ for all $A\in S(R)$ {\bf then} \\
\>\>\>\> \hspace{2mm} \=
$t_C.(P.t_i.A):=\bot$;\\
\> \> {\bf end} \\
\> {\bf end} \\
{\bf end}

\end{tabbing}
}
\hspace{4mm}
}
\vspace*{-1mm}

\end{small}
\caption{Evaluating relational algebra operations on WSDs.}
\label{fig:relop_alg}
\end{figure*}

%%% Local Variables: 
%%% mode: latex
%%% TeX-master: 569
%%% TeX-master: "569"
%%% End: 

\nop{ % moved for ICDE
\begin{example}
\em
Consider for example a query $\sigma_{A=1}(R) \cup \sigma_{B=2}(R)$.
If we first compute $\sigma_{A=1}(R)$, we must not replace $R$ by
$\sigma_{A=1}(R)$, otherwise $R$ will not be available for the computation of
$\sigma_{B=2}(R)$. On the other hand, if $\sigma_{A=1}(R)$ is stored in a
separate WSD, the connection between worlds of $R$ and the selection
$\sigma_{A=1}$ is lost and we can again not compute
$\sigma_{A=1}(R) \cup \sigma_{B=2}(R)$.
\punto
\end{example}
}

We say that a relation $P$ is a copy of another relation $R$ in a WSD if $R$
and $P$ have the same tuples in every world represented by
the WSD.
For a component $C$, an attribute $R.t.A_i$ of $C$ and a new attribute
$P.t.B$, the function \textsf{ext} extends $C$ by a new column $P.t.B$ that is
a copy of $R.t.A_i$:
\begin{align*}
  \textsf{ext} (C,A_i,B) := \{&(A_1:a_1,\ldots,A_n:a_n,B:a_i)\mid\\
  &(A_1:a_1,\ldots,A_n:a_n)\in C\}
\end{align*}
Then $\textsf{copy}(R,P)$ executes $C := \textsf{ext}(C, R.t_i.A,
P.t_i.A)$ for each component $C$ and each $R.t_i.A \in {\cal S}(C)$.

The implementation of some operations requires the composition of components.
Let $C_1$ and $C_2$ be two components with schemata $(A_1,\ldots,A_k,Pr)$, and \\
$(B_1,\ldots,B_l,Pr)$, respectively. Then the composition of $C_1$ and $C_2$ is defined as:
\begin{align*}
  \textsf{compose} &(C_1,C_2) :=\\
  \{(&A_1:a_1,\ldots,A_k:a_k,B_1:b_1,\ldots,B_l:b_l,\\
  & Pr: p_1 \cdot p_2)\mid\\
  &(A_1:a_1,\ldots,A_k:a_k, Pr: p_1)\in C_1,\\
  &(B_1:b_1,\ldots,B_l:b_l, Pr: p_2)\in C_2
  \}
\end{align*}

In the non-probabilistic case the composition of components is simply the relational product of the two components.

Figure~\ref{fig:relop_alg} presents implementations of the relational algebra operations
selection (of the form $\sigma_{A \theta c}$ or $\sigma_{A \theta B}$,
where $A$ and $B$ are attributes, $c$ is a constant, and $\theta$ is a
comparison operation, $=$, $\neq$, $<$, $\le$, $>$, or $\ge$), projection, relational
product and union on WSDs.  In each case, the
input WSD is {\em extended}\/ by the result of the operation.

\begin{figure*}[t!]
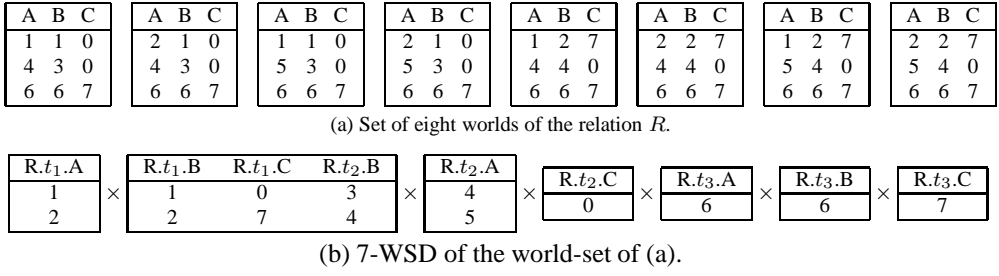

\begin{center}
{\footnotesize
%    \subfigure[Set of eight worlds of the relation $R$]
%    {
%     \label{fig:wsd1}
%\nop{ % removed for ICDE     
     \begin{tabular}{|c@{\extracolsep{2mm}}c@{\extracolsep{2mm}}c|}
    \hline
    A & B & C\\\hline
    1 & 1 & 0\\
    4 & 3 & 0\\
    6 & 6 & 7\\\hline
  \end{tabular}
  \hspace*{.5em}
  \begin{tabular}{|c@{\extracolsep{2mm}}c@{\extracolsep{2mm}}c|}
    \hline
    A & B & C\\\hline
    2 & 1 & 0\\
    4 & 3 & 0\\
    6 & 6 & 7\\\hline
  \end{tabular}
  \hspace*{.5em}
  \begin{tabular}{|c@{\extracolsep{2mm}}c@{\extracolsep{2mm}}c|}
    \hline
    A & B & C\\\hline
    1 & 1 & 0\\
    5 & 3 & 0\\
    6 & 6 & 7\\\hline
  \end{tabular}
  \hspace*{.5em}
  \begin{tabular}{|c@{\extracolsep{2mm}}c@{\extracolsep{2mm}}c|}
    \hline
    A & B & C\\\hline
    2 & 1 & 0\\
    5 & 3 & 0\\
    6 & 6 & 7\\\hline
  \end{tabular}
  \hspace*{.5em}
  \begin{tabular}{|c@{\extracolsep{2mm}}c@{\extracolsep{2mm}}c|}
    \hline
    A & B & C\\\hline
    1 & 2 & 7\\
    4 & 4 & 0\\
    6 & 6 & 7\\\hline
  \end{tabular}
  \hspace*{.5em}
  \begin{tabular}{|c@{\extracolsep{2mm}}c@{\extracolsep{2mm}}c|}
    \hline
    A & B & C\\\hline
    2 & 2 & 7\\
    4 & 4 & 0\\
    6 & 6 & 7\\\hline
  \end{tabular}
  \hspace*{.5em}
  \begin{tabular}{|c@{\extracolsep{2mm}}c@{\extracolsep{2mm}}c|}
    \hline
    A & B & C\\\hline
    1 & 2 & 7\\
    5 & 4 & 0\\
    6 & 6 & 7\\\hline
  \end{tabular}
  \hspace*{.5em}
  \begin{tabular}{|c@{\extracolsep{2mm}}c@{\extracolsep{2mm}}c|}
    \hline
    A & B & C\\\hline
    2 & 2 & 7\\
    5 & 4 & 0\\
    6 & 6 & 7\\\hline
  \end{tabular}
%}

\smallskip

(a) Set of eight worlds of the relation $R$.

\medskip
%} %end nop
%\subfigure[7-WSD of the world-set of Figure~\ref{fig:wsd1}]
%{
\begin{tabular}{|c|}
\hline
R.$t_1$.A \\
\hline
1 \\
2 \\
\hline
\end{tabular}
$\times$
  \begin{tabular}{|ccc|}
\hline
R.$t_1$.B & R.$t_1$.C & R.$t_2$.B \\
\hline
1 & 0 & 3\\
2 & 7 & 4\\
\hline
\end{tabular}
$\times$
\begin{tabular}{|c|}
\hline
R.$t_2$.A \\
\hline
4 \\
5 \\
\hline
\end{tabular}
$\times$
\begin{tabular}{|c|}
\hline
R.$t_2$.C \\
\hline
0 \\
\hline
\end{tabular}
$\times$
\begin{tabular}{|c|}
\hline
R.$t_3$.A \\
\hline
6 \\
\hline
\end{tabular}
$\times$
\begin{tabular}{|c|}
\hline
R.$t_3$.B \\
\hline
6 \\
\hline
\end{tabular}
$\times$
\begin{tabular}{|c|}
\hline
R.$t_3$.C \\
\hline
7 \\
\hline
\end{tabular}
}

\smallskip

% removed for ICDE
(b) 7-WSD of the world-set of (a).
%}%end footnotesize
\end{center}
\vspace*{-1em}
%\caption{7-WSD representing a a set of 8 worlds.}
\caption{World-set and its decomposition.}
\label{fig:wsd-running}
\end{figure*}

%%% Local Variables: 
%%% mode: latex
%%% TeX-master: main
%%% TeX-master: "main"
%%% End: 

\begin{figure*}[t!]
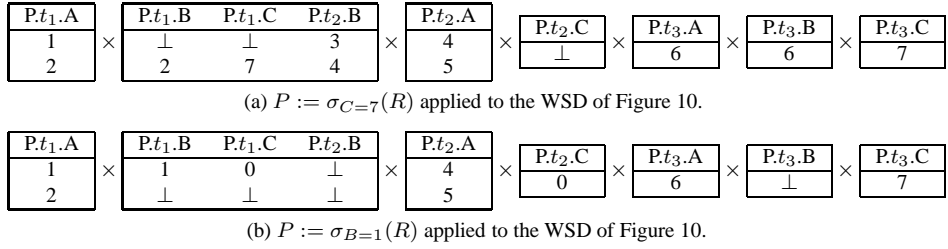

\begin{center}
{\footnotesize
\begin{tabular}{|c|}
\hline
P.$t_1$.A \\
\hline
1 \\
2 \\
\hline
\end{tabular}
$\times$
\begin{tabular}{|ccc|}
\hline
P.$t_1$.B  & P.$t_1$.C & P.$t_2$.B\\
\hline
$\bot$ & $\bot$ & 3 \\
2      & 7      & 4 \\
\hline
\end{tabular}
$\times$
\begin{tabular}{|c|}
\hline
P.$t_2$.A \\
\hline
4 \\
5 \\
\hline
\end{tabular}
$\times$
\begin{tabular}{|c|}
\hline
P.$t_2$.C \\
\hline
$\bot$ \\
\hline
\end{tabular}
$\times$
\begin{tabular}{|c|}
\hline
P.$t_3$.A \\
\hline
6 \\
\hline
\end{tabular}
$\times$
\begin{tabular}{|c|}
\hline
P.$t_3$.B \\
\hline
6 \\
\hline
\end{tabular}
$\times$
\begin{tabular}{|c|}
\hline
P.$t_3$.C \\
\hline
7 \\
\hline
\end{tabular}

\smallskip

%(a) $P:=\sigma_{C=7}(R)$ applied to the WSD of Figure~\ref{fig:wsd-running}~(b).
% ICDE version
(a) $P:=\sigma_{C=7}(R)$ applied to the WSD of Figure~\ref{fig:wsd-running}.
\medskip

\begin{tabular}{|c|}
\hline
P.$t_1$.A \\
\hline
1 \\
2 \\
\hline
\end{tabular}
$\times$
\begin{tabular}{|ccc|}
\hline
P.$t_1$.B  & P.$t_1$.C & P.$t_2$.B\\
\hline
1      & 0      & $\bot$ \\
$\bot$ & $\bot$ & $\bot$ \\
\hline
\end{tabular}
$\times$
\begin{tabular}{|c|}
\hline
P.$t_2$.A \\
\hline
4 \\
5 \\
\hline
\end{tabular}
$\times$
\begin{tabular}{|c|}
\hline
P.$t_2$.C \\
\hline
0 \\
\hline
\end{tabular}
$\times$
\begin{tabular}{|c|}
\hline
P.$t_3$.A \\
\hline
6 \\
\hline
\end{tabular}
$\times$
\begin{tabular}{|c|}
\hline
P.$t_3$.B \\
\hline
$\bot$ \\
\hline
\end{tabular}
$\times$
\begin{tabular}{|c|}
\hline
P.$t_3$.C \\
\hline
7 \\
\hline
\end{tabular}

\smallskip

%(b) $P:=\sigma_{B=1}(R)$ applied to the WSD of Figure~\ref{fig:wsd-running}~(b).
% ICDE version
(b) $P:=\sigma_{B=1}(R)$ applied to the WSD of Figure~\ref{fig:wsd-running}.

}%end footnotesize
\end{center}
%\vspace{-1em}

% \caption{Selections $P:=\sigma_{C=7}(R)$ and $P:=\sigma_{B=1}(R)$ with $R$ from Figure~\ref{fig:wsd-running}~(b).}
% ICDE version
\caption{Selections $P:=\sigma_{C=7}(R)$ and $P:=\sigma_{B=1}(R)$ with $R$ from Figure~\ref{fig:wsd-running}.}
\label{fig:constsel}
\end{figure*}

Let us now have a closer look at the evaluation of relational algebra
operations on WSDs.
 For this, we use as running example the set of eight worlds over the relation $R$ of
 Figure~\ref{fig:wsd-running}~(a) and its maximal 7-WSD of Figure~\ref{fig:wsd-running}~(b).
 The second component (from the left) of the
 WSD spans over several tuples and attributes and each of the remaining six
 components refer to one tuple and one attribute. The first tuple of the second
 component of the WSD of Figure~\ref{fig:wsd-running} contains the values
 for $R.t_1.B$, $R.t_1.C$, and $R.t_2.B$, i.e.\ some but not all of the
 attributes of the first and second tuple of $R^{\cal A}$, for all worlds
 ${\cal A}$. 
Because of space limitations and our attempt to keep the WSDs readable, we
consistently show in the following examples only the WSDs of the result
relations.

\medskip

%\paragraph*{Selection with condition $A \theta c$}

\noindent
{\underline{\bf Selection with condition $A \theta c$}}.
In order to compute a selection $P := \sigma_{A \theta c}(R)$, we first
compute a copy $P$ of relation $R$ and subsequently drop tuples of $P$ that do
not match the selection condition.

Dropping tuples is a fairly subtle operation, since
tuples can spread over several components and a component can define values for more than one tuple.

Thus a selection must not delete tuples from component relations, but should
mark fields as belonging to deleted tuples using the special value $\bot$. To
evaluate $\sigma_{A \theta c}(R)$, our selection algorithm of
Figure~\ref{fig:relop_alg} checks for each tuple
% \/\footnote{
%   The tuples over $P$ can be different -- even their number may vary -- in
%   each of the possible worlds over $P$. However, since world-set relations,
%   and therefore their decompositions, reserve a slot for the same $|P|_{max}$
%   tuples in each world, and just some worlds may have some of these tuples
%   marked as invalid, we may just as well refer to a tuple of $P$ as an object
%   that has a different {\em value}\/, and in some cases $\bot$, in each
%   possible world.}
$t_i$ in the relation $P$ and $t_C$ in component $C$ with attribute
$P.t_i.A$ whether $t_C.(P.t_i.A) \theta c$. In the negative case the
tuple $P.t_i$ is marked as deleted in all worlds that take values from
$t_C$. For that, $t_C.(P.t_i.A)$ is assigned value $\bot$, and all
other attributes $P.t_i.A'$ of $C$ referring to the same tuple $t_i$
of $P$ are assigned value $\bot$ in $t_C$,
(cf.\ the algorithm \textsf{propagate-$\bot$} of Figure~\ref{fig:propagation}).
This assures that if we later project away the
attribute $A$ of $P$, we do not erroneously ``reintroduce'' tuple $P.t_i$ into
worlds that take values from $t_C$.

% removed for ICDE
 \begin{figure}[h]
 \small
 \framebox[0.45\textwidth]{
 \hspace{2mm}
 \parbox{0cm}{
 \begin{tabbing}
 {\bf algorithm} \textsf{propagate-$\bot$}($C$: component)
 $\quad$\\
 {\bf begin} \\
 \hspace{2mm} \=
    {\bf for each} $t_C \in C$ and $P.t_i.A\in\mathcal{S}(C)$ {\bf do} \\
 \> \hspace{2mm} \=
  {\bf if} $t_C.(P.t_i.A) = \bot$ {\bf then} \\
 \> \> \hspace{2em} \=
    {\bf for each} $A'$ such that $P.t_i.A' \in S(C)$ {\bf do}\\
 \> \> \> \hspace{2mm} \=
         $t_C.(P.t_i.A') := \bot$; \\
 {\bf end}
 \end{tabbing}
 }}
% \vspace{-2mm}
 
 \caption{Propagating $\bot$-values.}
% \vspace{-4mm}
 \label{fig:propagation}
 \end{figure}

\begin{example}
  \em
  Figure~\ref{fig:constsel} shows the answers to $\sigma_{C=7}(R)$ and
  $\sigma_{B=1}(R)$.  Note that the resulting WSDs should contain both the
  query answer $P$ and the original relation $R$, but due to space limitations
  we only show the representation of $P$.  One can observe that for both
  results in Figure~\ref{fig:constsel} we obtain worlds of different sizes.
  For example the worlds that take values from the first tuple of the second
  component relation in Figure~\ref{fig:constsel}~(a) do not have a tuple
  $t_1$, while the worlds that take values from the second tuple of that
  component relation contain $t_1$.  \punto
\end{example}

% rewritten for ICDE
% \begin{example}
%   \em The answer $P$ to $\sigma_{C=7}(R)$ is represented by the WSD of
%   Figure~\ref{fig:constsel}~(a). Figure~\ref{fig:constsel}~(b) shows the
%   result of query $\sigma_{B=1}(R)$.  Note that the resulting WSDs should
%   contain both the query answer $P$ and the original relation $R$, but due to
%   space limitations we only show the representation of $P$.  One can observe
%   that for both results in Figure~\ref{fig:constsel} we may obtain worlds of
%   different sizes.  For example the worlds that take values from the first
%   tuple of the second component relation in Figure~\ref{fig:constsel}~(a) do
%   not have a tuple $t_1$, while the worlds that take values from the second
%   tuple of that component relation contain $t_1$.  \punto
% \end{example}

\begin{figure*}[t!]
{\footnotesize
\begin{center}
% end nop

%\subfigure[5-WSD after filtering all tuples using condition $A = B$]%
%{
%\label{fig:wsd7}
\begin{tabular}{|ccccc|}
\hline
P.$t_1$.A & P.$t_1$.B  & P.$t_1$.C & P.$t_2$.A & P.$t_2$.B\\
\hline
1      & 1      & 0      & $\bot$ & $\bot$\\
$\bot$ & $\bot$ & $\bot$ & $\bot$ & $\bot$\\
$\bot$ & $\bot$ & $\bot$ & 4 & 4\\
2      & 2      & 7      & 4 & 4\\
2      & 2      & 7      & $\bot$ & $\bot$\\
\hline
\end{tabular}
$\times$
\begin{tabular}{|c|}
\hline
P.$t_2$.C \\
\hline
0 \\
\hline
\end{tabular}
$\times$
\begin{tabular}{|cc|}
\hline
P.$t_3$.A & P.$t_3$.B\\
\hline
6 & 6\\
\hline
\end{tabular}
$\times$
\begin{tabular}{|c|}
\hline
P.$t_3$.C \\
\hline
7 \\
\hline
\end{tabular}
%}

\smallskip

% removed for ICDE
%(c) 5-WSD after filtering all tuples using condition $A = B$
\end{center}
}
\vspace*{-5mm}
% ICDE version
\caption{$P=\sigma_{A=B}(R)$ with $R$ from Figure~\ref{fig:wsd-running}.}
%\caption{$P=\sigma_{A=B}(R)$ with $R$ from Figure~\ref{fig:wsd-running}~(b).}
\label{fig:binary_sel_example}
\end{figure*}

%%% Local Variables: 
%%% mode: latex
%%% TeX-master: "paper"
%%% End: 

\begin{figure*}[t!]
{\footnotesize
\begin{center}
\begin{tabular}{|c|}
\hline
R.$t_1$.A \\
\hline
1 \\
2 \\
\hline
\end{tabular}
$\times$
  \begin{tabular}{|cc|}
\hline
R.$t_1$.B & R.$t_2$.A \\
\hline
3 & 5 \\
4 & 6 \\
\hline
\end{tabular}
$\times$
\begin{tabular}{|c|}
\hline
R.$t_2$.B \\
\hline
7 \\
8 \\
\hline
\end{tabular}
$\times$
\begin{tabular}{|c|}
\hline
S.$t_1$.C \\
\hline
a \\
b \\
\hline
\end{tabular}
$\times$
  \begin{tabular}{|cc|}
\hline
S.$t_1$.D & S.$t_2$.C \\
\hline
c & e \\
d & f \\
\hline
\end{tabular}
$\times$
\begin{tabular}{|c|}
\hline
S.$t_2$.D \\
\hline
g \\
h \\
\hline
\end{tabular}

\smallskip

(a) WSD of two relations $R$ and $S$.

\medskip

\begin{tabular}{|c@{\extracolsep{1mm}}c|}
\hline
$t_{11}$.A & $t_{12}$.A \\
\hline
1 & 1 \\
2 & 2 \\
\hline
\end{tabular}
$\times$
  \begin{tabular}{|c@{\extracolsep{1mm}}c@{\extracolsep{1mm}}c@{\extracolsep{1mm}}c|}
\hline
$t_{11}$.B & $t_{12}$.B & $t_{21}$.A & $t_{22}$.A \\
\hline
3 & 3 & 5 & 5 \\
4 & 4 & 6 & 6 \\
\hline
\end{tabular}
$\times$
\begin{tabular}{|c@{\extracolsep{1mm}}c|}
\hline
$t_{21}$.B & $t_{22}$.B \\
\hline
7 & 7 \\
8 & 8 \\
\hline
\end{tabular}
$\times$
\begin{tabular}{|c@{\extracolsep{1mm}}c|}
\hline
$t_{11}$.C & $t_{21}$.C \\
\hline
a & a \\
b & b \\
\hline
\end{tabular}
$\times$
  \begin{tabular}{|c@{\extracolsep{1mm}}c@{\extracolsep{1mm}}c@{\extracolsep{1mm}}c|}
\hline
$t_{11}$.D & $t_{21}$.D & $t_{12}$.C & $t_{22}$.C \\
\hline
c & c & e & e \\
d & d & f & f \\
\hline
\end{tabular}
$\times$
\begin{tabular}{|c@{\extracolsep{1mm}}c|}
\hline
$t_{12}$.D & $t_{22}$.D \\
\hline
g & g \\
h & h \\
\hline
\end{tabular}

\smallskip

(b) WSD of their product $R \times S$.
\end{center}
}

%\vspace{-5mm}

\caption{The product operation $R \times S$.}
\label{fig:prod_example}
\end{figure*}

%%% Local Variables: 
%%% mode: latex
%%% TeX-master: "paper"
%%% End: 

%\paragraph*{Selection with condition $A \theta B$}

\noindent
{\underline{\bf Selection with condition $A \theta B$}}.
The main added difficulty of selections with conditions $A \theta B$ as
compared to selections with conditions $A \theta c$ is that it creates dependencies between two attributes of
a tuple, which do not necessarily reside in the same component. 

As the current decomposition may not
capture exactly the combinations of values satisfying the join condition, components that have values for $A$ and $B$ of the same tuple are composed.  After the composition phase,
the selection algorithm follows the pattern of the selection with constant.

% rewritten for ICDE
% The 7-WSD of Figure~\ref{fig:wsd-running} has no component containing
% values for both the attributes $A$ and $B$ of any tuple of the decomposed
% worlds. Therefore, the test of the join condition for each of these tuples has
% to span over several components containing values for $A$ and $B$.
% Additionally, the join condition may exclude some possible combinations of the
% values from different components. Therefore, the current decomposition may not
% capture exactly the combinations of values satisfying the join condition. It
% may then be necessary to merge the components that have values for $A$ and $B$
% within a same tuple of the decomposed worlds.  After the composition phase,
% the selection algorithm follows the pattern of the selection with constant.

\begin{example}
  \em
  Consider the query $\sigma_{A=B}(R)$, where $R$ is represented
  by the 7-WSD of Figure~\ref{fig:wsd-running}.
  Figure~\ref{fig:binary_sel_example} shows the query answer, which is
  a 4-WSD that represents five worlds, where one world has three
  tuples, three worlds have two tuples each, and one world has one
  tuple.
\nop{ % changed for ICDE
Figure~\ref{fig:binary_sel_example}~(a)
shows the 6-WSD obtained from the one of
Figure~\ref{fig:wsd-running}~(b)
after  tuple $t_1$ has been processed (requiring the composition of the first
and second component of the WSD of Figure~\ref{fig:wsd-running}~(b)).
Further applying the selection to tuple $t_2$ yields the 5-WSD of
Figure~\ref{fig:binary_sel_example}~(b). Finally, processing tuple $t_3$ leads
to the composition of the second and the third components, as shown in
Figure~\ref{fig:binary_sel_example}~(c).  This 4-WSD represents five worlds,
where one world has three tuples, three worlds have two tuples each, and one
world has one tuple.
}
\punto
\end{example}

%\paragraph*{Projection}
%

\noindent
{\underline{\bf Product}}.
The product $T := R \times S$ of two relations $R$ and $S$, which have
disjunct attribute sets and are represented by a WSD requires
that the product relation $T$ extends a component $C$ with
$|S|_{max}$ (respectively $|R|_{max}$) copies of each column of $C$ with values of
$R$ (respectively $S$).  Additionally, the $i$th ($j$th) copy is named
$T.t_{ij}.A$ if the original has name $R.t_i.A$ or $S.t_j.A$.

\begin{example}
  \em
  Figure~\ref{fig:prod_example}~(b) shows the WSD for the product of
  relations $R$ and $S$ represented by the WSD of
  Figure~\ref{fig:prod_example}~(a). To save space, the relations $R$ and $S$
  have been removed from Figure~\ref{fig:prod_example}~(b), and attribute
  names do not show the relation name ``$T$''.  \punto
\end{example}

\noindent
{\underline{\bf Projection}}.  A projection $P=\pi_{U}(R)$ on an attribute set
$U$ of a relation $R$ represented by the WSD $\mathcal{C}$ is translated into
(1) the extension of $\mathcal{C}$ with the copy $P$ of $R$, and (2)
projections on the components of $\mathcal{C}$, where all component attributes
that do not refer to attributes of $P$ in $U$ are discarded. Before removing
attributes, however, we need to propagate $\bot$-values, as discussed in the
following example.

\begin{example}
  \em
  Consider the 3-WSD of Figure~\ref{fig:wsd8}~(a) representing a set of
  two worlds for $R$, where one world contains only the tuple $t_1$ and the
  other contains only the tuple $t_2$. Let $P'$ represent the first two
  components of $R$, which contain all values for the attribute $A$ in both
  tuples. The relation $P'$ is not the answer to $\pi_A(R)$, because it
  encodes one world with \textit{both} tuples, and the information from the
  third component of $R$ that only one tuple appears in each world is lost.
  To compute the correct answer, we progressively (1) compose the components
  referring to the same tuple (in this case all three components), (2)
  propagate $\bot$-values within the same tuple, and (3) project away the
  irrelevant attributes. The correct answer $P$ is given in
  Figure~\ref{fig:wsd8}~(b).
	\punto
\end{example}
% \vspace*{-4mm}

\begin{figure}[h!]
  {\footnotesize
	 	\begin{tabular}{c@{\extracolsep{0.8cm}}c}
	    \begin{tabular}{|@{~}c@{~}|}
	      \hline
	      R.$t_1$.A \\
	      \hline
	      a\\
	      \hline
	    \end{tabular}
	    $\times$
	    \begin{tabular}{|@{~}c@{~}|}
	      \hline
	      R.$t_2$.A \\
	      \hline
	      b\\
	      \hline
	    \end{tabular}
	    $\times$
	    \begin{tabular}{|@{~}c@{~}c@{~}|}
	      \hline
	      R.$t_1$.B & R.$t_2$.B\\
	      \hline
	      c & $\bot$\\
	      $\bot$ & d\\
	      \hline
	    \end{tabular}
	&
	%\hfill%
			\begin{tabular}{|@{~}c@{~}c@{~}|}
	      \hline
	      P.$t_1$.A & P.$t_2$.A\\
	      \hline
	      a & $\bot$\\
	      $\bot$ & b\\
	      \hline
	    \end{tabular}
		\\\\
		(a) WSD for R.
		&
		(a) WSD for P.
		\end{tabular}
}
\centering
% \vspace*{-2mm}

\caption{Projection $P:=\pi_{A}(R)$.}
% \vspace{-2mm}
\label{fig:wsd8}
\end{figure}

% removed for ICDE
%\nop{
The algorithm for projection is given in Figure~\ref{fig:relop_alg}. For each
tuple $t_i$, attribute $A$ in the projection list, and attribute $B$ not in
the projection list, the algorithm first propagates the $\bot$-values of
$P.t_i.B$ of component $C'$ to $P.t_i.A$ of component $C$. If $C$ and $C'$
are the same, the propagation is done locally within the component.
Otherwise, $C$ and $C'$ are merged before the propagation. Note that the
propagation is only needed if some tuples of $C'$ have at $\bot$-value for $t_i$.B.
This procedure is performed until no other components $C$ and $C'$ exist
that satisfy the above criteria. After the propagation phase, the attributes
not in the projection list are dropped from all remaining components.
%}

%% \begin{remark}
%% \em
%% Note that the correctness of projection depends on the following {\em
%%   invariant}\/ of our operator implementations.
%% For a component $C$ and any tuple $t_C \in C$, if there is a
%% field $t_C.(R.t.A)$ such that $t_C.(R.t.A) = \bot$, then for all $B$ such that
%% $R.t.B$ is an attribute of $C$, $t_C.(R.t.B) := \bot$.
%% \punto
%% \label{remark:invariant}
%% \end{remark}

%\paragraph*{Product}

%\paragraph*{Union, Difference, and Renaming}

\noindent
{\underline{\bf Union}}.
The algorithm for computing the union $T := R
\cup S$ of two relations $R$ and $S$ works similarly to that for the product.
Each component $C$ containing values of $R$ or $S$ is extended such that in
each world of $C$ all values of $R$ and $S$ become also values of $T$.

\noindent
{\underline{\bf Renaming}}.
The operation $\delta_{A \rightarrow A'}(R)$ renames attribute $A$ of
relation $R$ to $A'$ by renaming all attributes $R.t.A$ in a component $C$ to $R.t.A'$.

\noindent
{\underline{\bf Difference}}.
To compute the difference operation $P := R-S$ we scan and compose components of the two relations $R$ and $S$. For the worlds where a tuple $t$ from $R$ matches some tuple from $S$, we place $\bot$-values to denote that $t$ is not in these worlds of $P$; otherwise $t$ becomes a tuple of $P$.
The difference is by far the least efficient operation to implement, as it can lead to the composition of all components in the WSD.

%--------------------------------------------------------

%\nop{ % removed for ICDE
\section{Efficient Query Evaluation on UWSDTs}
\label{sec:alg-scalable}

The algorithms for computing the relational operations on WSDs
presented in Section~\ref{sec:queries} can be easily adapted to
UWSDTs. To do this, we follow closely the mapping of WSDs, represented
as sets of components $\mathcal{C}$, to equivalent UWSDTs, represented
by a triple ($F$,$C$,$W$) and at least one template relation
$R^0$:
\begin{itemize}
\item Consider a component $K$ of WSD $\mathcal{C}$ having an
  attribute $R.t.A$ with a value $v$. In the equivalent UWSDT, this
  value can be stored in the template relation $R^0$ if $v$ is the
  only value of $R.t.A$, or in the component $C$ otherwise. In the
  latter case, the template $R^0$ contains the placeholder $R.t.A$ in
  the tuple $t$. In addition, in the mapping relation $F$ there is an
  entry with the placeholder $R.t.A$ and a component identifier $c$,
  and $C$ contains a tuple formed by $R.t.A$, the value $v$ and a
  world identifier $w$.
%\begin{align}
%  &K\in\mathcal{C}\wedge R.t.A\in\mathcal{S}(K) \wedge v\in \pi_{t.A}(K) \Leftrightarrow \exists (c,w)\in W \nonumber\\
%  &(t\in R^0\wedge (R.t.A,c)\in F \wedge(R.t.A,w,v)\in C) \label{eq:mapping}
%\end{align}
  
\item Worlds of different sizes are represented in WSDs by allowing
  $\bot$ values in components, and in UWSDTs by allowing for a same
  placeholder different amount of values in different worlds.

\end{itemize}

%Recall that all four relations of a WSDT have fixed arities
%independent of the amount of worlds in the world-set, whereas the
%relations of a WSD have arities dependent on the represented
%world-set.

Any relational query is rewritten in our framework to a sequence of SQL
queries, except for the projection and selection with join conditions, where
the fixpoint computations are encoded as recursive PL/SQL programs. In all
cases, the size of the rewriting is linear in the size of the input query.
%Due to space limitations, we only give our efficient implementation of the
%selection with constant in Figure~\ref{fig:sel-uwsdt}.
Figure~\ref{fig:sel-uwsdt} shows the implementation of the
selection with constant on UWSDTs.

 \begin{figure}[pthb]
 \begin{small}
 \framebox[.5\textwidth]{
 \hspace{4mm}
 \parbox{9cm}{
 {\bf algorithm} select[$A \theta c$]
 $\quad$ // compute $P := \sigma_{A \theta c} R$ \\
 {\bf begin} \\
  \hspace*{.5em} 1. $P^0 := \sigma_{A\theta c\vee A=?}R^0$; \\
  \hspace*{.5em} 2. $F := F \cup \{(P.t.B,k) \mid (R.t.B,k) \in F, t\in P^0\}$; \\
  \hspace*{.5em} 3. $C := C \cup \{(P.t.B,w,v) \mid (R.t.B,w,v) \in C, t\in P^0,$ \\
  \hspace*{7em}     $(B=A\Rightarrow v\theta c)\}$; \\
  \hspace*{.5em} // Remove incomplete world tuples \\
% (see Remark~\ref{remark:invariant})\\
  \hspace*{.5em} 4. $C :=  C - \{(P.t.X,w,v) \in C \mid (P.t.X,k),(P.t.Y,k)\in F,$ \\
  \hspace*{7em}   $t\in P^0, X\not=Y, \not\exists v': (P.t.Y,w,v')\in C\}$; \\
  \hspace*{.5em} 5. $F :=  F - \{(P.t.B,k)\mid (P.t.B,k) \in F$,\\
  \hspace*{7em} $\not\exists w,v: (P.t.B,w,v)\in C\}$; \\
  \hspace*{.5em} 6. $P^0 :=  P^0 - \{t\mid t\in P^0, \not\exists B,a: (P.t.B,a)\in F\}$;\\
 {\bf end}
 }
 }
 \end{small}
 % \vspace{-2mm}

 \caption{Evaluating $P := \sigma_{A\theta c}(R)$ on UWSDTs.}
 % \vspace{-2mm}
 \label{fig:sel-uwsdt}
 \end{figure}
 
 In contrast to some algorithms of Figure~\ref{fig:relop_alg}, for UWSDTs we
 do not create a copy $P$ of $R$ at the beginning, but rather compute directly
 $P$ from $R$ using standard relational algebra operators. The template $P^0$
 is initially the set of tuples of $R^0$ that satisfy the selection condition,
 or have a placeholder `?' for the attribute $A$ (line 1). We extend the
 mapping relation $F$ with the placeholders of $P^0$ (line 2), and the
 component relation $C$ with the values of these placeholders, where the
 values of placeholders $P.t.A$ for the attribute $A$ must satisfy the
 selection condition (line 3). If a placeholder $P.t.A$ has no value
 satisfying the selection condition, then $t$ is removed from $P^0$ (line 6)
 and all placeholders of $t$ are removed from $F$ (line 5) together with their
 values from $C$ (line 4).
 
 Many of the standard query optimization techniques are also applicable in our
 context. For our experiments reported in Section~\ref{sec:experiments}, we
 performed the following optimizations on the sequences of SQL statements
 obtained as rewritings. For the evaluation of a query involving join, we
 merge the product and the selections with join conditions and distribute
 projections and selections to the operands. When evaluating a query involving
 several selections and projections on the same relation, we again merge these
 operators and perform the steps of the algorithm of
 Figure~\ref{fig:sel-uwsdt} only once. We further tuned the query evaluation
 by employing indices and materializing often used temporary results.

%} % end nop

%% \begin{example}
%%   \em To evaluate $P := \pi_{\textsf{A,C,D}}(\sigma_{\textsf{A}>0\vee
%%     \textsf{D}=0}R)$, we need the following minor extensions to the
%%   algorithm of Figure~\ref{fig:sel-uwsdt}. First, the selection
%%   condition becomes our condition disjunct $\textsf{A}>0\vee
%%   \textsf{D}=0$ augmented with checks on whether $A$ or $D$ contain
%%   placeholders. Second, we incorporate the projection in the
%%   computation of $P^0$ (line 1), add to $F$ only the placeholders of
%%   $P^0$ for the attributes in the projection list (line 2), and add to
%%   $C$ only the values of these placeholders (line 3).
%%  \punto
%% \end{example}

%%% Local Variables: 
%%% mode: latex
%%% TeX-master: main
%%% TeX-master: "main"
%%% End: 
\section{Confidence Computation in Probabilistic WSDs}
\label{sec:prob-wsd}
Section~\ref{sec:queries} discussed algorithms for evaluating relational algebra queries on top of WSDs. Since we consider queries that transform worlds, the algorithms were independent of whether or not probabilities were stored with the data.
A different class of queries are ones that compute confidence of tuples.
The {\em confidence} of a tuple $t$ in the result of a query $Q$ is defined as the sum of the probabilities of the worlds that contain $t$ in the answer to $Q$. Clearly, iterating over all possible worlds is infeasible. We therefore adopt an approach where we only iterate over the local worlds of the relevant components.

\begin{figure}[h]
	\small
	\framebox[0.45\textwidth]{
		\hspace{2mm}
		\parbox{4cm}{
			\begin{tabbing}
				// compute the confidence of tuple $t$ \\
				{\bf algorithm} conf(t)\\
				{\bf begin} \\
				\hspace{2mm} \=
							$c:=0$;\\
				\> {\bf let} $t_1,\ldots,t_k$ be the tuple ids\\
				\>$\quad$ that match $t$ in some world;\\
				\>	{\bf let} $C_1,\ldots,C_n$ be the components \\
				\> $\quad$ for the fields of $t_1,\ldots,t_k$;\\
				\>	{\bf let} $C:=\textsf{compose}(C_1,\ldots,C_n)$;\\
				\>  {\bf for each} $t_C$ in $C$ {\bf do} {\bf begin} \\
				\>\hspace{2mm} \=
						{\bf if} $t = (t_C.(t_i.A_1),\ldots,t_C.(t_i.A_m))$ for some $i$\\
				\>\>{\bf then} $c:=c+t_C.Pr$;\\ 
				\>{\bf end}\\
				\> return $c$;\\
				{\bf end}
			\end{tabbing}
		}
	}
	% \vspace{-2mm}
	
	\caption{Computing confidence of possible tuples.}
	\label{fig:t-conf}
	% \vspace{-4mm}
\end{figure}

Figure~\ref{fig:t-conf} shows an algorithm for computing the confidence of tuple $t$ of schema $(A_1,\ldots,A_m)$. It first finds those tuple ids $t_1,\ldots,t_k$ that match the given tuple $t$ in some world and composes all components defining fields of those tuple ids into one component $C$.
A world that contains $t$ is thus obtained whenever we select a local world from $C$ that makes the value of some tuple id $t_i$, $1\le i\le m$, equal to $t$. Fixing a local world in $C$ defines a set of possible worlds - the ones that share the values specified by the selected local world. The probability of this set of worlds is given in the $Pr$ field of the local world. Since the local worlds of a component define non-overlapping sets of worlds, to compute the confidence of $t$ we need to sum up the probabilities of those local worlds that define $t$. 

Note that the algorithms for computing tuple confidence in~\cite{dalvi04efficient} rely heavily on the fact that input tuples are independent. Tuple confidence is computed during the evaluation of the query in question to avoid having to store intermediate results. This restricts the supported types of queries and the query plans that can be used. In probabilistic WSDs on the other hand, the query evaluation can be completely decoupled from confidence computation, since the latter form a strong representation system. For the same reason we need no independence assumptions about the input data.
	
The algorithm of Figure~\ref{fig:t-conf} does not explore possible independence between tuples.
One can design a better approach in the following way.
In a probabilistic WSD each component id corresponds to an independent random variable, whose possible outcomes are the local worlds of the component.
We will call a {\em world-set descriptor (ws-descriptor)} a set
\[
	\{(C_1,L_1),\ldots,(C_n,L_n)\}
\]
where $C_i$ is a component id, $L_i$ is a local world id of $C_i$, and no two elements $(C_i,L_i),(C_j,L_j)$ of the set exist with $C_i=C_j$ and $L_i\neq L_j$. A ws-descriptor defines, as its name suggests, a set of possible worlds, whose probability can be computed as the product of the probabilities of the selected local worlds:
\[
	P(\{(C_1,L_1),\ldots,(C_n,L_n)\}) = \prod\limits_{i=1}^n P(C_i,L_i)
\]
A ws-descriptor that specifies a local world for each component id of a probabilistic WSD corresponds to a single world.
%
%\begin{example}
%The world-set descriptor $\{(C_1,1),(C_2,1)\}$ for the WSD of Figure~\ref{fig:dalvi-wsd} defines the set of worlds that contain tuples $s_1$ and $s_2$ and may or may not contain $t_1$. The probability of these worlds is computed as $0.8\cdot0.5=0.4$. 
%\punto
%\end{example}
%
For computing tuple confidence we need to also consider sets of ws-descriptors. A ws-descriptor set defines a set of possible worlds - the union of the worlds defined by each descriptor in the set.
Given a fixed tuple $t$ and a probabilistic world-set decomposition ${\cal W}$ representing the answer $R$ to query $Q$, we compute a ws-descriptor set $D$ for the worlds containing $t$ in the following way. Let $t_i$ be a tuple id of $R$ and $C_{i_1},\ldots,C_{i_j}$ be the components of ${\cal W}$ that define fields of $R.t_i$. If the value of $t_i$ is $t$ when we fix the local world of $C_{i_k}$ to be $L_{i_k}$ for $1\le k\le j$, respectively, then $D$ contains the ws-descriptor $\{(C_{i_1},L_{i_1}),\ldots,(C_{i_j},L_{i_j})\}$. The confidence of $t$ is then computed as the probability of the worlds defined by $D$.
Computing tuple confidence can be reduced to computing the probability of a formula in disjunctive normal form, which is known to have \#P complexity. 
This follows
from the mutual reducibility of the problem of computing
the probability of the union of the (possibly overlapping)
world-sets represented by a set of ws-descriptors and of the
\#P-complete problem of counting the number of satisfying
assignments of Boolean formulas in disjunctive normal form.
Indeed, we can encode a set of $k$ ws-descriptors
$\{\{(C_{i_1},L_{i_1}),\ldots,(C_{i_j},L_{i_j})\}\}, 1\le i\le k$
as a formula
$\bigvee\limits_{1\le i\le k}(C_{i_1}=L_{i_1}\land\ldots\land C_{i_j}=L_{i_j})$.
Different optimization techniques exist for computing the probability of a boolean formula, such as variable elimination and Monte Carlo approximations~\cite{re2007topk}.

\begin{remark}
	\em
	The U-relations of \cite{ajko08} associate each possible combination of values with a ws-descriptor. In WSDs and UWSDTs on the other hand a combination of values is associated with a single pair of component and local world id. Thus WSDs form a special case of U-relations with dependency vectors of size one.
	\punto
\end{remark}

We next consider the operator possible that computes the tuples appearing in at least one world of the world-set. Formally, if $R$ is a relation name and ${\bf A}$ - a world-set, the operator possible is defined as:
\begin{align*}
  \textsf{possible} &(R)({\bf A}) := \{t\mid {\cal A} \in {\bf A}, t\in R^{\cal A}\}
\end{align*}

\begin{figure}[h]
	\small
	\framebox[0.45\textwidth]{
		\hspace{2mm}
		\parbox{4cm}{
			\begin{tabbing}
				// compute $P := \textsf{possible}(R)$ \\
				{\bf algorithm} possible\\
				{\bf begin} \\
				\hspace{2mm} \=
					$P:=\emptyset$;\\
				\>   {\bf for each} $1 \le i \le |R|_{max}$ {\bf do} {\bf begin} \\
				\> \hspace{2mm} \=
				      {\bf let} $C_1,\ldots,C_k$ be the components for\\
				\>\>	\quad the fields of $R.t_i$; \\
				\>\>	{\bf let} $C:=\textsf{compose}(C_1,\ldots,C_k)$; \\ 
				\>\>	add $\pi_{R.t_i.A_1,\ldots,R.t_i.A_m}(\sigma_{\bigwedge_j R.t_i.A_j\neq\bot}(C))$ to $P$;\\
				\>{\bf end}\\
				{\bf end}
			\end{tabbing}
		}
	}
	% \vspace{-2mm}
	
	\caption{Computing possible tuples.}
	\label{fig:poss}
	% \vspace{-4mm}
\end{figure}

Figure~\ref{fig:poss} shows an algorithm for computing possible tuples in the non-probabilistic case. For each tuple id $t_i$ for $R$ we compose the components defining fields of $t_i$ to obtain the possible values for $t_i$.

\begin{figure}[h]
\small
	\framebox[0.45\textwidth]{
		\hspace{2mm}
		\parbox{4cm}{
			\begin{tabbing}
				// compute $P := \textsf{possible}^p(R)$ \\
				{\bf algorithm} possible$^p$\\
				{\bf begin} \\
				\hspace{2mm} \=
					$P:=\emptyset$;\\
				\> {\bf for each} distinct $t$ in $\textsf{possible}(R)$ {\bf do} {\bf begin} \\
				\> \hspace{2mm} \=
					add $(t, conf(t))$ to $P$;\\
				{\bf end}
			\end{tabbing}
		}
	}
	% \vspace{-2mm}
	
	\caption{Computing possible tuples together with their confidence.}
	\label{fig:p-possible}
	% \vspace{-4mm}
\end{figure}

In the probabilistic case the operator possible can be extended to compute the confidence of the possible tuples. To do that, we compute the confidence of each tuple $t$, which is a possible answer to $Q$.
Figure~\ref{fig:p-possible} shows an algorithm implementing the operator possible in the probabilistic case that computes the possible tuples together with their confidence. For computing the confidence $conf(t)$ of tuple $t$ we can plug in any exact or approximate algorithm, e.g.\ the one from Figure~\ref{fig:t-conf}.

\begin{example}
\em
Consider the probabilistic WSD of Figure~\ref{fig:prob-wsd}, query $Q=\pi_{S}(R)$, and tuple $t=(185)$. Let $C_1$ denote the first component. This component represents the answer to the projection query. There are two tuple ids whose values match the given tuple $t$, and they are already defined in the same component $C_1$. To compute the confidence of $t$ we therefore need to sum up the probabilities of the first and second local world, obtaining $0.2 + 0.4=0.6$. The following table contains the possible tuples in the answer to $Q$ together with their confidence:
\begin{center}
	\begin{small}
	\begin{tabular}{c|c|c}
		$Q$ & S & conf\\\hline
		& 185 & 0.6 \\
		& 186 & 0.6 \\
		& 785 & 0.8 \\
		
	\end{tabular}
	\end{small}
\end{center}
\punto
\end{example}

%Finally, to compute the confidence of all tuples in the answer to query $Q$, we compute the confidence of each tuple $t$, which is a possible answer to $Q$.
%Figure~\ref{fig:conf} shows an algorithm implementing the operator possible in the probabilistic case that computes the possible tuples together with their confidence.
%
%\begin{example}
%Consider the probabilistic WSD of Figure~\ref{fig:prob-wsd} and the query $Q=\pi_{S}(R)$. Let $t=(185)$. If $C_1$ denotes the first component and $1,2,3$ are its local worlds, the world-set descriptor set defining the worlds that contain $t$ in the answer to $Q$ is then $\left\{\{(C_1,1)\},\{(C_1,2)\}\right\}$. Since the two descriptors define non-overlapping sets of worlds, their probability is computed simply as the sum $0.2 + 0.4=0.6$. The following table contains the possible tuples in the answer to $Q$ together with their confidence:
%\begin{center}
%	\begin{small}
%	\begin{tabular}{c|c|c}
%		$Q$ & S & conf\\\hline
%		& 185 & 0.6 \\
%		& 186 & 0.6 \\
%		& 785 & 0.8 \\
%		
%	\end{tabular}
%	\end{small}
%\end{center}
%\punto
%\end{example}

\section{Normalizing probabilistic WSDs}
The normalization of a WSD is the
process of finding an equivalent probabilistic WSD that takes the least space among all its equivalents.
Examples of not normalized WSDs are non-maximal WSDs or WSDs defining
invalid tuples (i.e., tuples that do not appear in any world). Note
that removing invalid tuples and maximizing world-set decompositions
can be performed in polynomial time~\cite{ako06worlds}.

Figure~\ref{fig:normalization} gives three algorithms that address these
normalization problems. The third algorithm scans for identical tuples in a component and compresses them into one by summing up their probabilities.

\begin{figure}[h]
	\small
	\framebox[0.45\textwidth]{
	\hspace{2mm}
	\parbox{4cm}{
	\begin{tabbing}
	{\bf algorithm} remove\_invalid\_tuples\\
	{\bf begin} \\
	\hspace{2mm} \=
	   {\bf for each} $1 \le i \le |P|_{max}$ and $A\in\mathcal{S}(P)$ {\bf do} {\bf begin} \\
	\> \hspace{2mm} \=
	      {\bf let} $C$ be the component of $P.t_i.A$; \\
	\> \> {\bf if} $\pi_{P.t_i.A} = \{\bot\}$ {\bf then}\\
	\> \> \hspace{2mm} \=
	            {\bf for each} $B\in\mathcal{S}(P)$ {\bf do begin} \\
	\> \> \> \hspace{2mm} \=
	            let $C'$ be the component of $P.t_i.B$; \\
	\> \> \> \> project away $P.t_i.B$ from $C'$; \\
	\> \> \> {\bf end} \\
	\> {\bf end} \\
	{\bf end}\\
	\\
	{\bf algorithm} decompose\\
	{\bf begin} \\
	\hspace{2mm} \=
	   {\bf while} no fixpoint is reached {\bf do} {\bf begin} \\
	\> \hspace{2mm} \=
	      {\bf let} $C$ be a component such that\\
	\>\> \quad $C=\textsf{compose}(C_1, C_2)$;\\
	\> \> replace $C$ by $C_1$, $C_2$;\\
	\> {\bf end} \\
	{\bf end}\\
	\\
	{\bf algorithm} compress\\
	{\bf begin} \\
	\hspace{2mm} \=
	   {\bf while} no fixpoint is reached {\bf do} {\bf begin} \\
	\> \> {\bf let} $C$ be a component, $w_1,w_2\in C$ such that\\
	\> \> \quad $w_1.A=w_2.A$ for all $A\in S(C),A\neq Pr$;\\
	\> \> {\bf let} $w$ be a tuple such that $w.Pr:=w_1.Pr+w_2.Pr$,\\
	\> \> \quad $w.A:=w_1.A$ for all $A\in S(C),A\neq Pr$;\\
	\> \> replace $w_1,w_2$ in $C$ by $w$;\\
	\> {\bf end}\\
	{\bf end}
	\end{tabbing}
	}
	}
	% \vspace{-2mm}
	\caption{Algorithms for WSD normalization.}
	\label{fig:normalization}
	% \vspace{-4mm}
\end{figure}

\begin{example}
  \em
  The WSD of Figure~\ref{fig:constsel}~(a) has only $\bot$-values for
  $P.t_2.C$. This means that the tuple $t_2$ of $P$ is absent (or invalid) in
  all worlds and can be removed. The equivalent WSD of Figure~\ref{fig:simpl}
  shows the result of this operation. Similar simplifications apply to the WSD
  of Figure~\ref{fig:constsel}~(b), where tuples $t_2$ and $t_3$ are invalid.
  \punto
\end{example}

\begin{figure}[h!]
	\centering
	{\footnotesize
		\begin{tabular}{|@{~}c@{~}|}
			\hline
			P.$t_1$.A \\
			\hline
			1 \\
			2 \\
			\hline
		\end{tabular}
		$\times$
		\begin{tabular}{|@{~}c@{~}c@{~}|}
			\hline
			P.$t_1$.B & P.$t_1$.C\\
			\hline
			$\bot$ & $\bot$\\
			2 & 7\\
			\hline
		\end{tabular}
		$\times$
		\begin{tabular}{|@{~}c@{~}|}
			\hline
			P.$t_3$.A\\
			\hline
			6 \\
			\hline
		\end{tabular}
		$\times$
		\begin{tabular}{|@{~}c@{~}|}
			\hline
			P.$t_3$.B \\
			\hline
			6 \\
			\hline
		\end{tabular}
		$\times$
		\begin{tabular}{|@{~}c@{~}|}
			\hline
			P.$t_3$.C \\
			\hline
			 7 \\
			\hline
		\end{tabular}
	}
% \vspace{-2mm}

\caption{Normalization of WSD of Figure~\ref{fig:constsel}~(a).}
% \vspace{-6mm}
\label{fig:simpl}
\end{figure}

\begin{example}
  \em
  The 4-WSD of Figure~\ref{fig:binary_sel_example} admits the
  equivalent 5-WSD, where the third component is decomposed into two
  components. This non-maximality case cannot appear for UWSDTs,
  because all but the first component contain only one tuple and
  are stored in the template relation, where no component merging
  occurs.
  %Such a case of non-maximality is detected by the algorithm \textsf{decompose}.
%
\punto
\end{example}

\section{Experimental Evaluation}
\label{sec:experiments}

The literature knows a number of approaches to representing incomplete
information databases, but little work has been done so far on expressive yet
efficient representation systems.  An ideal representation system would allow a large set of possible worlds to be managed using only a small overhead in storage space
and query processing time when compared to a single world represented in a
conventional way. In the previous sections we presented the first step towards
this goal. 
%We introduced UWSDTs and studied the query processing problem in this context.
This section reports on experiments with a large census database with noise
represented as a UWSDT.

% from chr: redundant with next paragraph:
%
%, by  addressing the following questions: What is the space
%overhead of the additional relations used to represent incomplete
%information?  Is the performance of query evaluation on UWSDTs
%comparable to the simpler case of querying one of their worlds?
%
%How big are UWSDTs after chasing a set of real-life dependencies on
%them?
%
%

%
%The only parameter changes are for shared memory and cache:
%\textsf{shared\_buffers} for Postgres and \textsf{shmall} for Linux
%were set to 160 MB, and the effective cache size of Postgres to 800
%MB.
%-----------------------------
\noindent
{\underline{\bf Setting}}. The experiments were conducted on a 3GHz/\\
2GB Pentium machine running Linux 2.6.8 and PostgreSQL 8.0.

% We sound like users if we say that
%
%The queries were applied via the psql interface of PostgreSQL.

\noindent
{\underline{\bf Datasets}}. The IPUMS 5\% census data (Integrated Public Use
Microdata Series, 1990)~\cite{IPUMS} used for the experiments is the publicly
available 5\% extract from the 1990 US census, consisting of 50
(exclusively) multiple-choice questions.
It is a relation with 50
attributes and 12491667 tuples (approx.\ 12.5 million). The size of this
relation stored in PostgreSQL is ca.\ 3 GB.
We also used excerpts representing the
first 0.1, 0.5, 1, 5, 7.5, and 10 million tuples.

\noindent
{\underline{\bf Adding Incompleteness}}. We added incompleteness as
follows.  First, we generated a large set of possible worlds by
introducing noise. After that, we cleaned the data by removing worlds
inconsistent with respect to a given set of dependencies. Both steps
are detailed next.

We introduced noise by replacing some values with or-sets\footnote{We consider
  it infeasible both to iterate over all worlds in secondary storage, or to
  compute UWSDT decompositions by comparing the worlds.}. We experimented with
different noise densities: 0.005\%, 0.01\%, 0.05\%, 0.1\%. For example, in the
0.1\% scenario one in 1000 fields is replaced by an or-set. The size of each
or-set was randomly chosen in the range $[2, min(8,size)]$, where $size$ is
the size of the domain of the respective attribute (with a measured average of
3.5 values per or-set). In one scenario we had far more than $2^{624449}$
worlds, where 624449 is the number of the introduced or-sets and 2 is the
minimal size of each or-set (cf.\@ Figure~\ref{fig:results-size}).

We then performed data cleaning using 12 equality generating dependencies, representing real-life constraints on the census data.
%as the ones of Figure~\ref{fig:dependencies}.
% These represent real-life constraints on the census data, such as citizens born in
% the USA are not immigrants.
%, and the second one requires that citizens
%who served in the second world war have done their military service.
Note that or-set relations are not expressive enough to represent the
cleaned data with dependencies.

To remove inconsistent worlds with respect to given dependencies, we
adapted the Chase technique~\cite{AHV95} to the context of UWSDTs. We
explain the Chase by an example. Consider the dependency WWII = 1
$\Rightarrow$ MILITARY != 4 that requires people who participated in
the second world war to have completed their military service.  Assume
now the dependency does not hold for a tuple $t$ in some world and let
$C_1$ and $C_2$ be the components defining $t$.WWII and $t$.MILITARY,
respectively. First, the Chase computes a component $C$ that defines
both $t$.WWII and $t$.MILITARY.  In case $C_1$ and $C_2$ are
different, they are replaced by a new component $C = C_1 \times C_2$;
otherwise, $C$ is $C_1$. The Chase removes then from $C$ all
inconsistent worlds $w$, i.e., worlds where $w$.WWII = 1 and
$w$.MILITARY = 4. Repeating these steps iteratively for each
dependency on a given UWSDT yields a UWSDT satisfying all
dependencies.

\nop{ % removed for ICDE
\begin{figure}[th]
  {\footnotesize
  \begin{tabular}{|r|llclr|}
    \hline
  1 & CITIZEN & $=$ 0 & $\Rightarrow$ & IMMIGR   & $=$ 0 \\
%   2 & FEB55   & $=$ 1 & $\Rightarrow$ & MILITARY & $!=$ 4 \\
%   3 & KOREAN  & $=$ 1 & $\Rightarrow$ & MILITARY & $!=$ 4 \\
%   4 & VIETNAM & $=$ 1 & $\Rightarrow$ & MILITARY & $!=$ 4 \\
  2 & WWII    & $=$ 1 & $\Rightarrow$ & MILITARY & $!=$ 4 \\
%   6 & MARITAL & $=$ 0 & $\Rightarrow$ & RSPOUSE  & $!=$ 6 \\  
%   7 & MARITAL & $=$ 0 & $\Rightarrow$ & RSPOUSE  & $!=$ 5 \\
%   8 & LANG1   & $=$ 2 & $\Rightarrow$ & ENGLISH  & $!=$ 4 \\  
%   9 & RPOB    & $=$ 52& $\Rightarrow$ & CITIZEN  & $!=$ 0 \\
%  10 & SCHOOL  & $=$ 0 & $\Rightarrow$ & KOREAN   & $!=$ 1 \\
%  11 & SCHOOL  & $=$ 0 & $\Rightarrow$ & FEB55    & $!=$ 1 \\
%  12 & SCHOOL  & $=$ 0 & $\Rightarrow$ & WWII     & $!=$ 1 \\
\hline
 \end{tabular}\centering
  \caption{Example dependencies for cleaning census data.}
  \label{fig:dependencies}
  }
\end{figure}
}
%\begin{figure}[htbp]
%  \centering
%  \includegraphics{time-chase}

%\vspace{-3mm}

%  \caption{Time for chasing the dependencies of Figure~\ref{fig:dependencies} on UWSDTs of various sizes and densities.}
%  \label{fig:time-chase}
%\end{figure}

\begin{figure}[t!]
  \centering
  \begin{center}
    {\footnotesize
  \begin{tabular}{l|l|lllll|}
         & Density        & 0.005\% & 0.01\%     & 0.05\%     & 0.1\%\\\hline
 Initial & \#comp         & 31117   & 62331      & 312730     & 624449\\\hline
After    & \#comp         & 30918   & 61791      & 309778     & 612956\\
chase    & \#comp$>$1     & 249     & 522        & 2843       & 10880 \\
         & $|C|$          & 108276  & 217013     & 1089359    & 2150935\\
         & $|R|$          & 12.5M   & 12.5M      & 12.5M      & 12.5M   \\\hline
After    & \#comp         & 702     & 1354       & 7368       & 14244     \\
$Q_1$    & \#comp$>$1     & 1       & 4          & 40         & 158       \\
         & $|C|$          & 1742    & 3625       & 19773      & 37870     \\
         & $|R|$          & 46600   & 46794      & 48465      & 50499     \\\hline
After    & \#comp         & 25      & 56         & 312        & 466       \\
$Q_2$    & \#comp$>$1     & 0       & 1          & 8          & 9         \\
         & $|C|$          & 93      & 269        & 1682       & 2277      \\
         & $|R|$          & 82995   & 83052      & 83357      & 83610     \\\hline
After    & \#comp         & 38      & 76         & 370        & 742       \\
$Q_3$    & \#comp$>$1     & 0       & 0          & 0          & 0         \\
         & $|C|$          & 89      & 202        & 1001       & 2009      \\
         & $|R|$          & 17912   & 17936      & 18161      & 18458     \\\hline
After    & \#comp         & 1574    & 3034       & 15776      & 30729     \\
$Q_4$    & \#comp$>$1     & 11      & 28         & 127        & 557       \\
         & $|C|$          & 4689    & 9292       & 48183      & 94409     \\
         & $|R|$          & 402345  & 402524     & 404043     & 405869    \\\hline
After    & \#comp         & 3       & 10         & 53         & 93        \\
$Q_5$    & \#comp$>$1     & 3       & 10         & 53         & 93        \\
         & $|C|$          & 1221    & 5263       & 33138      & 50780     \\
         & $|R|$          & 150604  & 173094     & 274116     & 393396    \\\hline
After    & \#comp         & 97      & 189        & 900        & 1888      \\
$Q_6$    & \#comp$>$1     & 0       & 0          & 0          & 0         \\
         & $|C|$          & 516     & 1041       & 4993       & 10182     \\
         & $|R|$          & 229534  & 230113     & 234335     & 239488    \\\hline
  \end{tabular}
}
\end{center}

% \vspace{-3mm}

  \caption{UWSDTs characteristics for 12.5M tuples.}
% \vspace{-4mm}
  \label{fig:results-size}
\end{figure}

%\vspace{-4mm}

\begin{figure}[htbp]
{\footnotesize
\framebox[.47\textwidth]{
\hspace{2mm}
\parbox{7cm}{
% \vspace{-1mm}
\begin{align*}
  Q_1 &:= \sigma_{\textsf{YEARSCH} = 17\wedge \textsf{CITIZEN} = 0}(R)\\
  Q_2 &:= \pi_{\textsf{POWSTATE,CITIZEN,IMMIGR}}(\sigma_{\textsf{CITIZEN}<>0\wedge \textsf{ENGLISH}>3}(R))\\
  Q_3 &:= \pi_{\textsf{POWSTATE,MARITAL,FERTIL}}(\sigma_{\textsf{POWSTATE}=\textsf{POB}}\\
  &\hspace*{1.5em} (\sigma_{\textsf{FERTIL}>4\wedge \textsf{MARITAL}=1}(R)))\\
  Q_4 &:= \sigma_{\textsf{FERTIL}=1\wedge (\textsf{RSPOUSE}=1\vee\textsf{RSPOUSE}=2)}(R) \\
  Q_5 &:= \delta_{\textsf{POWSTATE}\rightarrow P_1}(\sigma_{\textsf{POWSTATE}>50}(Q_2)) \bowtie_{P_1 = P_2}\\
  &\hspace*{1.5em} \delta_{\textsf{POWSTATE}\rightarrow P_2}(\sigma_{\textsf{POWSTATE}>50}(Q_3))\\
  Q_6 &:= \pi_{\textsf{POWSTATE,POB}}(\sigma_{\textsf{ENGLISH}=3}(R))
\end{align*}
}
}
}
% \vspace{-2mm}

\caption{Queries on IPUMS census data.}
 % \vspace{-4mm}
\label{fig:queries}
\end{figure}

Figure~\ref{fig:results-size} shows the effect of chasing our dependencies on
the 12.5 million tuples and varying placeholder density. As a result of
merging components, the number of components with more than one placeholder
(\#comp$>$1) grows linearly with the increase of placeholder density, reaching
about 1.7\% of the total number of components (\#comp) in the 0.1\% case. A
linear increase is witnessed also by the chasing time when the number of
tuples is also varied. 

%% Due to lack of space, we do not report further on experiments with our Chase
%% procedure.

%Figure~\ref{fig:datasize} shows that an increase of the or-set density
%has a logarithmic effect on the relative size increase of the UWSDT,
%when chasing the dependencies of Figure~\ref{fig:dependencies}. This
%can be explained by (1) the small probability to get or-sets within a
%same tuple, as well as by (2) the large number of equality-generating
%dependencies, which may only lead to component compositions in case
%(1) above. The single functional dependency can lead to composition of
%components spanning over several tuples and attributes, but this case
%is quite rare even for our maximum or-set probability of 5\%.
%The relative time increase for chasing our dependencies is shown in
%Figure~\ref{fig:time-chase} only for the 1\% IPUMS data.  Here, the
%chasing time on UWSDTs, generated under varying or-set probability, is
%considered relative to the chasing time on the UWSDT with the or-set
%probability zero, which corresponds exactly to the 1\% IPUMS data.

%-----------------------------
%Selection with constants is used by $Q_1$ to $Q_5$, selections with
%join conditions by $Q_3$ and $Q_5$, projections by $Q_2$, $Q_3$, and
%$Q_6$, and join by $Q_5$.

\begin{figure*}[htbp]
  \centering
    \includegraphics[scale=.69]{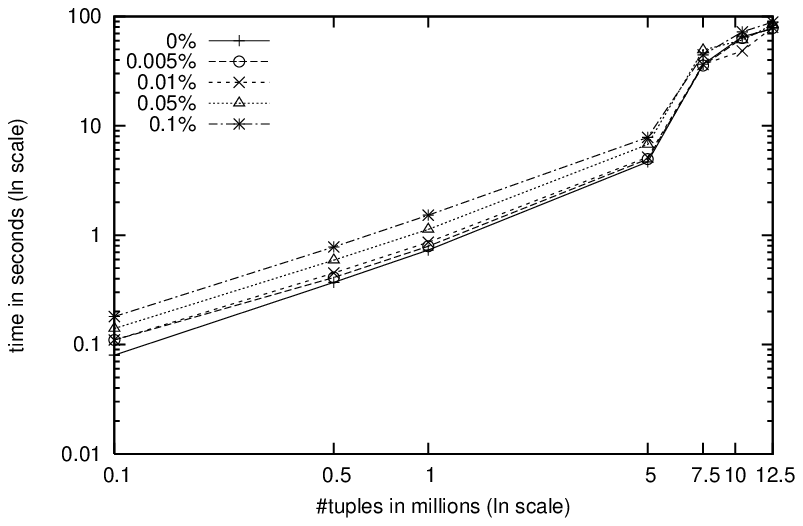}
    \includegraphics[scale=.69]{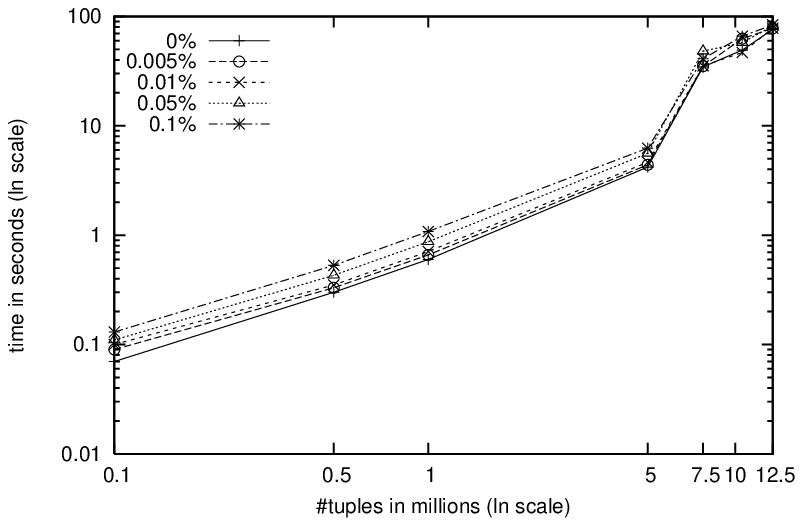}
    \includegraphics[scale=.69]{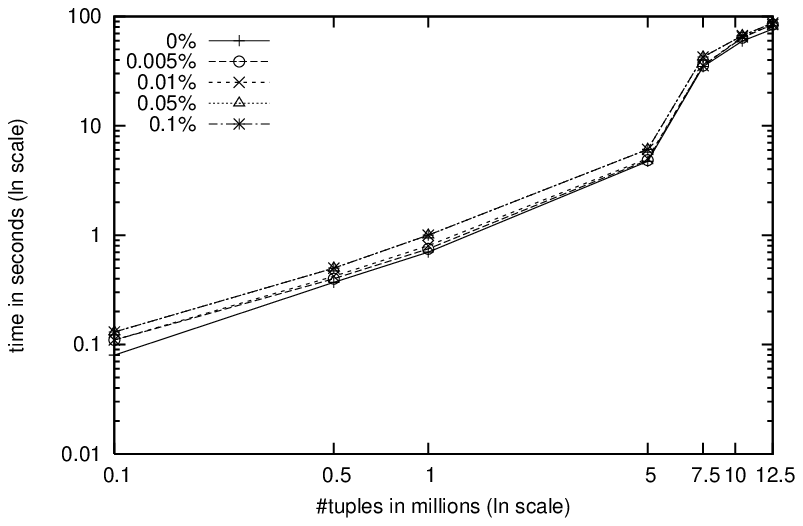}
  \hspace{1cm} (a) Query $Q_1$ \hspace{3.8cm} (b) Query $Q_2$ \hspace{3.8cm} (c) Query $Q_3$
    \includegraphics[scale=.69]{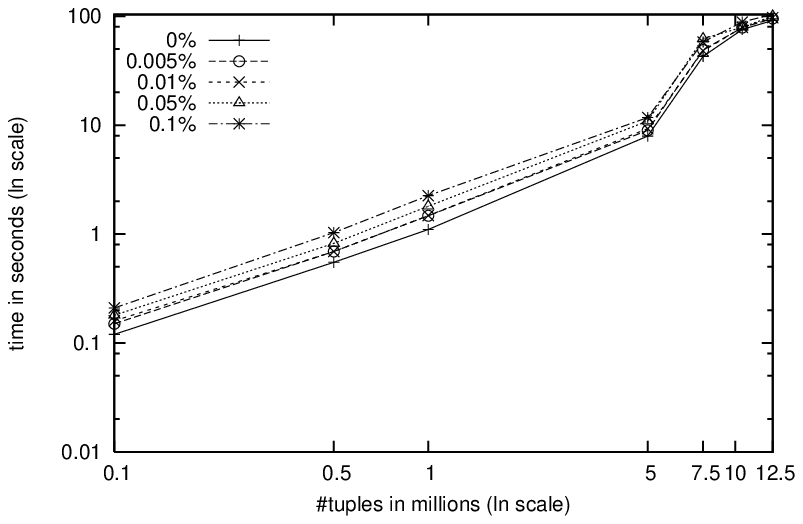}
    \includegraphics[scale=.69]{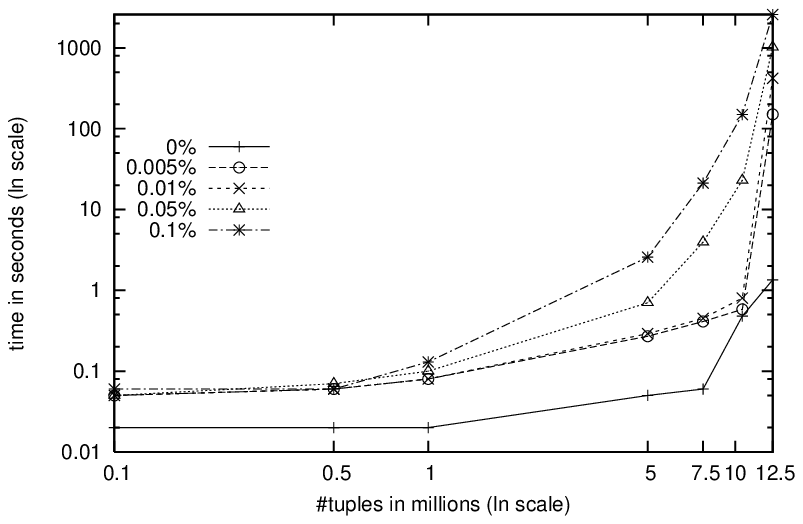}
    \includegraphics[scale=.69]{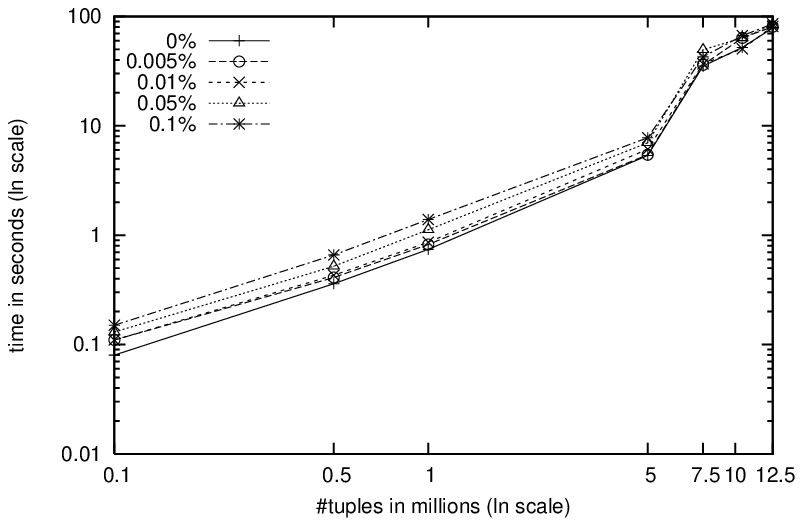}
  \hspace{1cm} (d) Query $Q_4$ \hspace{3.8cm} (e) Query $Q_5$ \hspace{3.8cm} (f) Query $Q_6$

  %\vspace{-2mm}

  \caption{The evaluation time for queries of Figure~\ref{fig:queries} on UWSDTs of various sizes and densities.}
  %\vspace{-2mm}
  \label{fig:results-querytime}
\end{figure*}

\noindent
{\underline{\bf Queries}}. Six queries were chosen to show the behavior of
relational operators combinations under varying selectivities (cf.\ 
Figure~\ref{fig:queries}). Query $Q_1$ returns the entries of US citizens with
PhD degree. The less selective query $Q_2$ returns the place of birth of US
citizens born outside the US that do not speak English well. Query $Q_3$
retrieves the entries of widows that have more than three children and live in
the state where they were born. The very unselective query $Q_4$ returns all
married persons having no children. Query $Q_5$ uses query $Q_2$ and $Q_3$ to
find all possible couples of widows with many children and foreigners with
limited English language proficiency in US states with IPUMS index greater
than 50 (i.e., eight `states', e.g., Washington, Wisconsin, Abroad). Finally,
query $Q_6$ retrieves the places of birth and work of persons speaking English
well.

Figure~\ref{fig:results-size} describes some characteristics of the answers to
these queries when applied on the cleaned 12.5M tuples of IPUMS data: the
total number of components (\#comp) and of components with more than one
placeholder (\#comp$>$1), the size of the component relation $C$, and the size
of the template relation $R$.  One can observe that the number of components
increases linearly with the placeholder density and that compared to chasing,
query evaluation leads to a much smaller amount of component merging.

Figure~\ref{fig:results-querytime} shows that all six queries admit
efficient and scalable evaluation on UWSDTs of different sizes and
placeholder densities. For accuracy, each query was run ten times, and
the median time for computing and storing the answer is reported. The
evaluation time for all queries but $Q_5$ on UWSDTs follows very
closely the evaluation time in the one-world case. The one-world case
corresponds to density 0\% in our diagrams, i.e., when no placeholders
are created in the template relation and consequently there are no
components. In this case, the original queries (that is, not the
rewritten ones) of Figure~\ref{fig:queries} were evaluated only on the
(complete) template relation.

An interesting issue is that all diagrams of
Figure~\ref{fig:results-querytime} show a substantial increase in the query
evaluation time for the 7.5M case.  As the jump appears also in the one-world
case, it suggests poor memory management of Postgres in the case of large
tables. We verified this statement by splitting the 12.5M table into chunks
smaller than 5M and running query $Q_1$ on those chunks to get partial
answers.  The final answer is represented then by the union of each UWSDT
relation from these partial answers.

Although the evaluation of join conditions on UWSDTs can require theoretically
exponential time (due to the composition of some components), our
experiments suggest that they behave well in practical cases, as illustrated
in Figures~\ref{fig:results-querytime}~(c) and (e) for queries $Q_3$ and $Q_5$
respectively. Note that the time reported for $Q_5$ does not include the time
to evaluate its subqueries $Q_2$ and $Q_3$.

In summary, our experiments show that UWSDTs behave very well in practice.  We
found that the size of UWSDTs obtained as query answers remains close to that
of one of their worlds. Furthermore, the processing time for queries on UWSDTs
is comparable to processing one world. The explanation for this is that in
practice there are rather few differences between the worlds. This keeps the
mapping and component relations relatively small and the lion's share of the
processing time is taken by the templates, whose sizes are about the same as
of a single world.

%%% Local Variables: 
%%% mode: latex
%%% TeX-master: main
%%% TeX-master: "main"
%%% End: 
\section{Application Scenarios}
\label{sec:applications}

Our approach is designed to cope with large sets of possible worlds, which
exhibit local dependencies and large commonalities. This data pattern can be
found in many applications. In addition to the census scenario used in Section
\ref{sec:experiments}, we next discuss two further application scenarios that
can profit from our approach. As for the census scenario, we consider it
infeasible both to iterate over all possible worlds in secondary storage, or
to compute UWSDT decompositions by comparing the worlds. Thus we also outline
how our UWSDTs can be efficiently computed.

%Inconsistent Databases
%An application scenario that lies closely to our approach is the management and querying of inconsistent databases.

\noindent
{\underline{\bf Inconsistent databases.}} A database is inconsistent if it
does not satisfy given integrity constraints. Sometimes, enforcing the
constraints is undesirable. One approach to manage such inconsistency is to
consider so-called {\em minimal repairs}, i.e., consistent instances of the
database obtained with a minimal number of changes \cite{ABC1999}. A repair
can therefore be viewed as a possible (consistent) world. The number of
possible minimal repairs of an inconsistent database may in general be
exponential; however, they substantially overlap. For that reason repairs can
be easily modeled with UWSDTs, where the consistent part of the database is
stored in template relations and the differences between the repairs in
components. Current work on inconsistent databases \cite{ABC1999} focuses on
finding {\em consistent query answers}, i.e., answers appearing in all
possible repairs (worlds). With our approach we can provide more than that, as
the answer to a query represents a set of possible worlds. In this way, we
preserve more information that can be further processed using querying or data
cleaning techniques.

%medicinenet

\noindent
{\underline{\bf Medical data.}} Another application scenario is modeling
information on medications, diseases, symptoms, and medical procedures, see,
e.g., \cite{medicinenet}. A particular characteristic of such
data is that it contains a big number of clusters of interdependent data. For
example, some medications can interact negatively and are not approved for
patients with some diseases. Particular medical procedures can be prescribed
for some diseases, while they are forbidden for others. In the large set of
possible worlds created by the complex interaction of medications,
diseases, procedures, and symptoms, a particular patient record can represent
one or a few possible worlds. Our approach can keep interdependent data within
components and independent data in separate components. One can ask then for
possible patient diagnostics, given an incompletely specified medical history
of the patient, or for commonly used medication for a given set of diseases.

In \cite{medicinenet} interdependencies of
medical data are modeled as links. A straightforward and efficient approach to wrap such
data in UWSDTs is to follow the links and create one component for all
interrelated values. Additionally, each different kind of information, like
medications, diseases, is stored in a separate template relation.

%%% Local Variables: 
%%% mode: latex
%%% TeX-master: 569
%%% TeX-master: "569"
%%% End: 

%\input{futurework}

\bibliographystyle{plain}
\bibliography{bibtex}

\end{document}